\documentclass[a4paper,british,11pt]{article}

\usepackage[T1]{fontenc}
\usepackage[utf8]{inputenc}
\usepackage[british]{babel}
\usepackage[
  rm={oldstyle=false,proportional=true},
  sf={oldstyle=false,proportional=true},
  tt={oldstyle=false,proportional=true}
]{cfr-lm}
\usepackage[babel]{microtype}

\usepackage{amsmath}
\usepackage{amsthm}

\usepackage{complexity}
\usepackage{xspace}

\usepackage{authblk}

\usepackage[misc]{ifsym}
\usepackage{orcidlink}

\usepackage{hyperref}
\hypersetup{
  hidelinks,
  bookmarksnumbered,
  pdftitle={Winning and losing NoGo is PSPACE-complete},
  pdfauthor={Kyle Burke, Alfie Davies, and Craig Tennenhouse}
}
\usepackage[capitalise,noabbrev]{cleveref}

\newtheorem{theorem}{Theorem}[section]
\newtheorem{lemma}[theorem]{Lemma}
\newtheorem{corollary}[theorem]{Corollary}

\AddToHook{env/lemma/begin}{\crefalias{theorem}{lemma}}
\AddToHook{env/corollary/begin}{\crefalias{theorem}{corollary}}
\AddToHook{env/remark/begin}{\crefalias{theorem}{remark}}

\usepackage{tikz}

\NewDocumentCommand{\GoBoardScaled}{m O{#1} O{1} m m +O{}}{%
  \begin{tikzpicture}[
    grid/.style={black!45, thin},
    white/.style={ball color=white, draw=black!35, very thin},
    black/.style={ball color=black},
    liberty/.style={fill=white, inner sep=1pt, font=\bfseries},
    signal/.style={->},
    signal label/.style={inner sep=1pt, font=\scriptsize},
    scale=#3,
    trim left=(board-left),
    trim right=(board-right)
  ]
    \coordinate (board-left) at (-0.5,0);
    \coordinate (board-right) at ({#1+0.5},0);
    \foreach \x in {0,...,#1} {
      \draw[grid] (\x,0) -- ++(0,#2);
    }
    \foreach \y in {0,...,#2} {
      \draw[grid] (0,\y) -- ++(#1,0);
    }
    \foreach \stone in {#4} {
      \shadedraw[white] \stone circle (4.9mm);
    }
    \foreach \stone in {#5} {
      \shade[black] \stone circle (4.9mm);
    }
    #6
  \end{tikzpicture}%
}

\newcommand*{\ruleset}[1]{\textsc{#1}}
\newcommand*{\rsNoGo}{\ruleset{NoGo}\xspace}
\newcommand*{\rsGNoGo}{\ruleset{Graph NoGo}\xspace}
\newcommand*{\rsBoundedCL}{\ruleset{Bounded Two-Player Constraint Logic}\xspace}
\newcommand*{\rsBCL}{\ruleset{B2CL}\xspace}
\newcommand*{\rsCAL}{\ruleset{Coins-are-Lava}\xspace}

\newcounter{authcount}
\NewDocumentCommand{\authDetails}{m m m O{\arabic{authcount}}}{%
  \stepcounter{authcount}%
  \author[#4]{%
    \mbox{#1\,$^{\IfBlankF{#2}{\textrm{\href{mailto:#2}{\Letter}}\,\,}%
      \IfBlankF{#3}{\raisebox{-0.2ex}{\orcidlink{#3}}\,}}$}%
  }%
}

\title{Winning and losing \rsNoGo is \PSPACE-complete}

\authDetails{Kyle Burke}{kburke@flsouthern.edu}{0000-0002-9222-8832}[1]
\authDetails{Alfie Davies\thanks{Supported by Leverhulme Trust Research
  Project Grant RPG-2024-182.}}
  {research@alfied.xyz}{0000-0002-4215-7343}[2]
\authDetails{Craig Tennenhouse}{ctennenhouse@une.edu}{0000-0002-0600-0963}[3]

\affil[1]{Florida Southern College, USA}
\affil[2]{Department of Computer Science, Durham University, UK}
\affil[3]{School of Mathematics and Data Science, University of New England, USA}

\date{}

\begin{document}

\maketitle

\begin{abstract}
  \noindent
  We resolve an open problem of Burke and Hearn \cite{BurkeHearn2018} and show
  that the combinatorial game \rsNoGo is \PSPACE-complete on grids. This is the
  usual geometry the game is played on, including at the tournament at a 2011
  Banff workshop. In addition to this result, we show that it is also
  \PSPACE-complete under mis\`ere play on grids.  We prove both of these
  results using reductions from \rsBoundedCL. Finally, we show that it is
  \PSPACE-complete under normal play on connected bipartite subcubic graphs.
\end{abstract}

\section{Introduction}

Combinatorial game theory studies two-player games of perfect and complete
information and no chance moves
\cite{LessonsInPlay:2007,WinningWays:2001,SiegelCGT:2013}. One such game is
\rsNoGo, a variant of \ruleset{Go} in which captures (including self-captures)
and passes are illegal. In 2005, John Moore described an early version of the
game on Sensei's Library under the name \ruleset{Anti-Atari
Go}~\cite{moore:anti-atari-go}. Neil McKay independently invented \rsNoGo in
2011~\cite[p.~396]{huggan.nowakowski:conjoined}.

We concisely describe the rules of \rsNoGo for completeness: Black and White
take turns placing a stone of their own colour onto an empty point of the
square grid. A \emph{group} is a connected component of the stones of one
colour. The \emph{liberties} of a group are the empty points adjacent to it. A
move is legal if it leaves every group (of either colour) with at least one
liberty. (In \ruleset{Go}, removing all liberties from a group \emph{captures}
it and removes its stones from the board.) Under normal play, a player loses
if they cannot make a legal move on their turn, whereas under mis\`ere play,
a player \emph{wins} if they cannot make a legal move on their turn.\footnote{
  \rsNoGo is playable at \url{https://kyleburke.info/DB/combGames/noGo.html}
  for normal play, and at
  \url{https://kyleburke.info/DB/combGames/noGoMisere.html} for mis\`ere play.
} The same rules define \rsGNoGo on arbitrary finite graphs, where the vertices
are the points of the board and the edges specify adjacency.

A fundamental problem within algorithmic combinatorial game theory is
\emph{winnability}: that is, deciding whether the player to move has a winning
strategy from a given position~\cite{AlgGameTheoryGONC3}. In this paper, we
analyse whether Black has a winning strategy playing first in normal and
mis\`ere play. Burke and Hearn report that the \NP-hardness of \rsGNoGo on
planar graphs was established at the 2011 Banff workshop, where a \rsNoGo
tournament was held~\cite[p.~408]{BurkeHearn2018}. They later proved that the
problem is \PSPACE-complete on planar graphs, and asked whether the same holds
on the square grid~\cite[Theorem~4 and Open Problem~1]{BurkeHearn2018}. This
open problem was highlighted again by Buchin, Hagedoorn, Kostitsyna, and van
Mulken after they proved \ruleset{Dots and Boxes} to be
\PSPACE-complete~\cite{buchin.ea:dots-and-boxes}. Our main result answers their
question affirmatively, on positions reachable through play starting from an
empty board. Alongside recent work on the complexity of mis\`ere versions of
well-known combinatorial games~\cite{2511.21888}, we also establish the same
hardness under mis\`ere play.

\begin{theorem}
  \label{thm:square-grid}
  \rsNoGo is \PSPACE-complete on positions reachable from an
  empty $n\times n$ square grid under both normal play and mis\`ere play.
\end{theorem}

Burke and Hearn first reduced \rsBoundedCL (\rsBCL) to planar \ruleset{Col},
and then planar \ruleset{Col} to planar \rsGNoGo~\cite{BurkeHearn2018}. For
\cref{thm:square-grid}, we realise the \rsBCL gadgets directly on the square
grid. We connect our gadgets with wires that are typically realised as chains
of White stones with two liberties, one in each gadget; if Black fills in one
of these liberties, then another Black move at the other would capture the
chain, regardless of how long the wire is (which is illegal).

Although it was not explicitly stated in Burke and Hearn's planar hardness
result for \rsGNoGo, an inspection of their gadget reduction demonstrates
hardness on planar graphs of maximum degree at most 11. \cref{thm:square-grid}
shows hardness on planar graphs of maximum degree at most 4. Does hardness
persist to graphs of maximum degree at most 3 (i.e.\ on subcubic graphs)? We
show that it does, although with no planarity restriction.

\begin{theorem}
  \label{thm:subcubic}
  \rsGNoGo is \PSPACE-complete under normal play on connected bipartite
  subcubic graphs, even for positions whose game trees are impartial.
\end{theorem}

In Moore's original post about \ruleset{Anti-Atari Go} (i.e.\ \rsNoGo) in 2005,
he observed that it ``plays much like the game of dots and
boxes''~\cite{moore:anti-atari-go}. This turned out to be remarkably prescient.
In the dual description of \ruleset{Dots and Boxes}, each box becomes a coin,
and each side a string; drawing a side of a box cuts that string. This
generalises to arbitrary (multi)graphs as \ruleset{Strings-and-Coins}, whose
normal play variant is \ruleset{Nimstring}. To be clear: we have vertices
called \emph{coins}, and the edges between them are called \emph{strings}, and
players take turns removing edges (i.e.\ cutting strings); if a player frees a
coin, then they play again.

To obtain \cref{thm:subcubic}, we reduce from a further minor variant called
\rsCAL. Here, players take turns cutting strings as before, except that a
player who cuts the last string attached to any coin immediately loses the game.  (The scores do not matter.)  
On simple graphs, this is the ``edge-delete game'' of Gallant, Gunther,
Hartnell, and Rall, who analysed the game on paths and graphs with special
symmetries~\cite{gallant.ea:edge-delete}. Demaine and Diomidova proved that
\rsCAL is \PSPACE-complete (in their general multigraph setting) in
\cite[Theorem~1]{demaine.diomidova:strings-and-coins}. We adapt their
construction to obtain connected bipartite instances of \rsCAL, which then
yield the subcubic instances asserted by \cref{thm:subcubic}. The basic idea is
to replace each coin by a path of stones forming one group, and each string by
a liberty shared between two groups (one White and one Black). Cutting the
final string attached to a coin is then equivalent to claiming the final
liberty of a group.

Given \cref{thm:subcubic}, a straightforward construction also allows us to
realise the very same impartial game trees after uniformly subdividing every
edge of our subcubic graphs $\ell\ge1$ times. This connects our results to the
C123 framework of Johnson et al.~\cite{johnson.ea:C123} for graph classes
defined by forbidden subgraphs. Its first condition (C1) concerns tractability
on classes of bounded treewidth. Our results here establish the two required
hardness conditions: hardness on subcubic graphs (C2), and hardness on their
uniform $\ell$-subdivisions for arbitrarily large $\ell$ (C3). It is then
immediate from \cite[proofs of Theorems~4 and~6]{johnson.ea:C123} that \rsGNoGo
is \PSPACE-complete on $H$-subgraph-free graphs for every fixed graph $H$ that
has a component other than a path or a subdivided claw. The cases not covered
by this argument are therefore only those in which $H$ is a disjoint union of
paths and subdivided claws.

\section{Normal play on the square grid}

In order to prove that \rsNoGo is \PSPACE-hard under normal play, we reduce
from \rsBCL.  When reducing from \rsBCL, it is more common to reduce from hard
instances built in \cite{DBLP:books/daglib/0023750} (as in
\cite{BurkeHearn2018,2511.21888}) than directly from a general \rsBCL position.
To that end, we need to build gadgets for the roles of VARIABLE, AND, OR,
SPLIT, and CHOICE, as well as a Goal and others to make the system work as a
whole.  The reason for this is that with those gadgets, \rsBCL grants us
planarity, necessary for \rsNoGo on a grid.  Although reductions of this sort
appear like reductions from formula games (\ruleset{Positive CNF} in
particular), they are actually sourced by
\rsBCL~\cite{DBLP:books/daglib/0023750}.

The five required gadgets have a formula-like interpretation.  VARIABLE must
pass on an active signal if one player (Black for us) plays first on it and
inactive if the other player (White) plays first.  The players are then forced
to play so that AND yields an active signal if and only if both inputs are
active.  OR yields an active signal if and only if at least one of the inputs
is active.  SPLIT yields two active outputs if the input is active, otherwise
both are inactive.  Finally, CHOICE yields two inactive outputs if the input is
inactive, otherwise Black chooses exactly one of the outputs to be active.

The general plan for our construction is that wires of white stone paths
surrounded by black stones will connect the gadgets.  If a wire is active, then
there will still be a liberty for that white path on the input side, allowing a
black stone to be played on the output end of the wire.  Black will be required
to play a certain number of stones per gadget; otherwise they will lose.  In
general, by being able to block the incoming wire(s), they will not have to
play stones that block off the outgoing wire(s).

\begin{figure}[htbp]
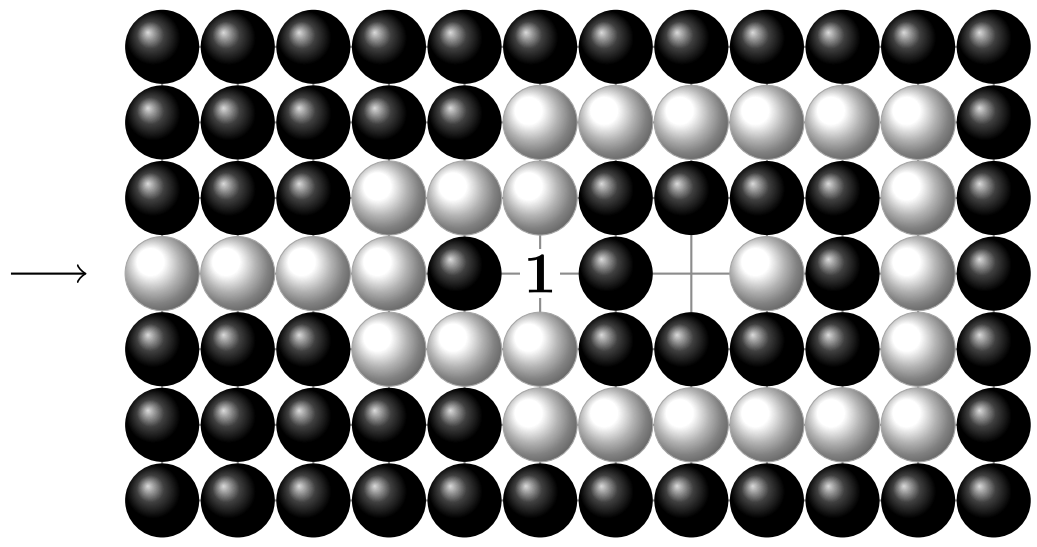

  \centering
  \GoBoardScaled{11}[6][.5]
    {(0,3), (1,3), (2,3), (3,3), (3,4), (4,4), (5,4), (5,5), (6,5), (7,5), (8,5), (9,5), (10,5), (10,4), (10,3), (10,2), (10,1), (9,1), (8,1), (7,1), (6,1), (5,1), (5,2), (4,2), (3,2),
    (8,3)}
    {(0,2), (1,2), (2,2), (2,1), (3,1), (4,1), (4,0), (5,0), (6,0), (7,0), (8,0), (9,0), (10, 0), (11,0), (11,1), (11,2), (11,3), (11,4), (11,5), (11,6), (10,6), (9,6), (8,6), (7,6), (6,6), (5,6), (4,6), (4,5), (3,5), (2,5), (2,4), (1,4), (0,4),
    (9,3), (9, 4), (8,4), (7,4), (6,4), (6,3), (6,2), (7,2), (8,2), (9,2),
    (4,3),
    (3,0), (2,0), (1,0), (0,0), (0,1), (1,1),
    (3,6), (2,6), (1,6), (0,6), (0,5), (1,5)}
    [
      \draw[signal] (-2,3) -- (-1,3);
      \node[liberty] at (5,3) {1};
    ]
  \caption{Goal gadget. The input enters through the left-hand side.}
  \label{fig:normalGoal}
\end{figure}

Black will win the game if they play the prescribed number of times in each
gadget and are able to play on the Goal gadget, which we describe in
\cref{fig:normalGoal}.  In that gadget there are two open spaces, but only
space 1 is playable, only by Black, and only if the input white wire has a
liberty on its other end.  Thus, Black can make that play only if the input to
this gadget is active.

\begin{figure}[htbp]
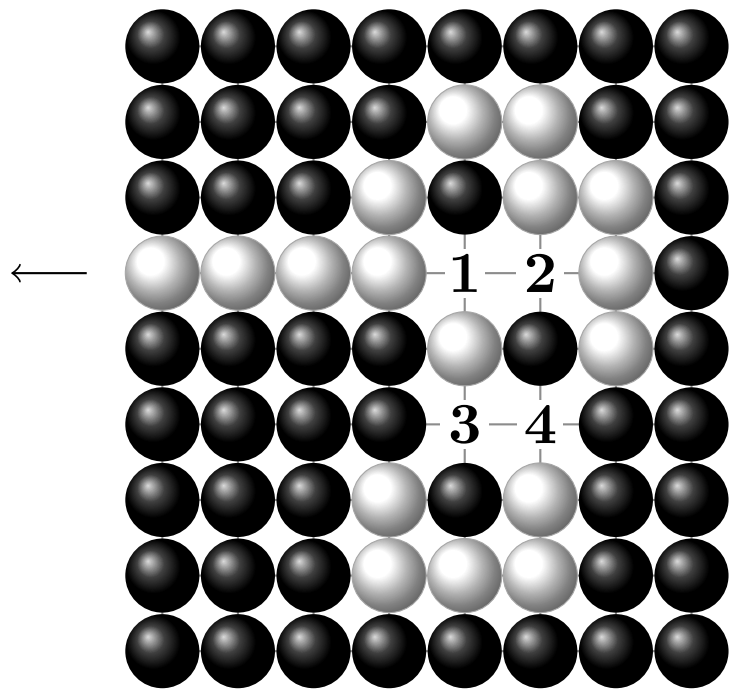

  \centering
  \GoBoardScaled{7}[8][.5]
    {(0,5), (1,5), (2,5), (3,5), (3,6),
    (4,4),
    (3,2), (3,1), (4,1), (5,1), (5,2),
    (6,4), (6,5), (6,6), (5,6), (5,7), (4,7)}
    {(0,4), (1,4), (2,4), (3,4),
    (3,3), (2,3), (2,2), (2,1), (2,0), (3,0), (4,0), (5,0), (6,0), (6,1), (6,2), (6,3), (7,3), (7,4), (7,5), (7,6), (7,7), (6,7), (6,8), (5,8), (4,8), (3,8), (3,7), (2,7), (2,6), (1,6), (0,6),
    (4,2), (5,4), (4,6),
    (0,0), (0,1), (0,2), (0,3), (1,3), (1,2), (1,1), (1,0),
    (7,0), (7,1), (7,2),
    (7,8),
    (1,7), (0,7), (0,8), (1,8), (2,8)}
    [
      \draw[signal] (-1,5) -- (-2,5);
      \node[liberty] at (4,5) {1};
      \node[liberty] at (5,5) {2};
      \node[liberty] at (4,3) {3};
      \node[liberty] at (5,3) {4};
    ]
  \caption{VARIABLE gadget.  The output signal exits on the left-hand side.}
  \label{fig:normalVariable}
\end{figure}

Next we describe the VARIABLE gadget, shown in \cref{fig:normalVariable}.
There are four open spaces (labelled 1, 2, 3, and 4) but only Black can play on
1 and 3 and only White can play on 2 and 4.  Additionally, only one of 1 and 2,
only one of 3 and 4, only one of 1 and 3, and only one of 2 and 4 can be
played.  Thus each player can play at most once on the gadget.  If Black plays
first, it will never benefit them to choose 1 over 3 because playing at 1
deactivates the output.  If White plays first, it is in their interest to play
on 4 so that Black's only play on the gadget can be at 1, deactivating the
output.

\begin{figure}[htbp]
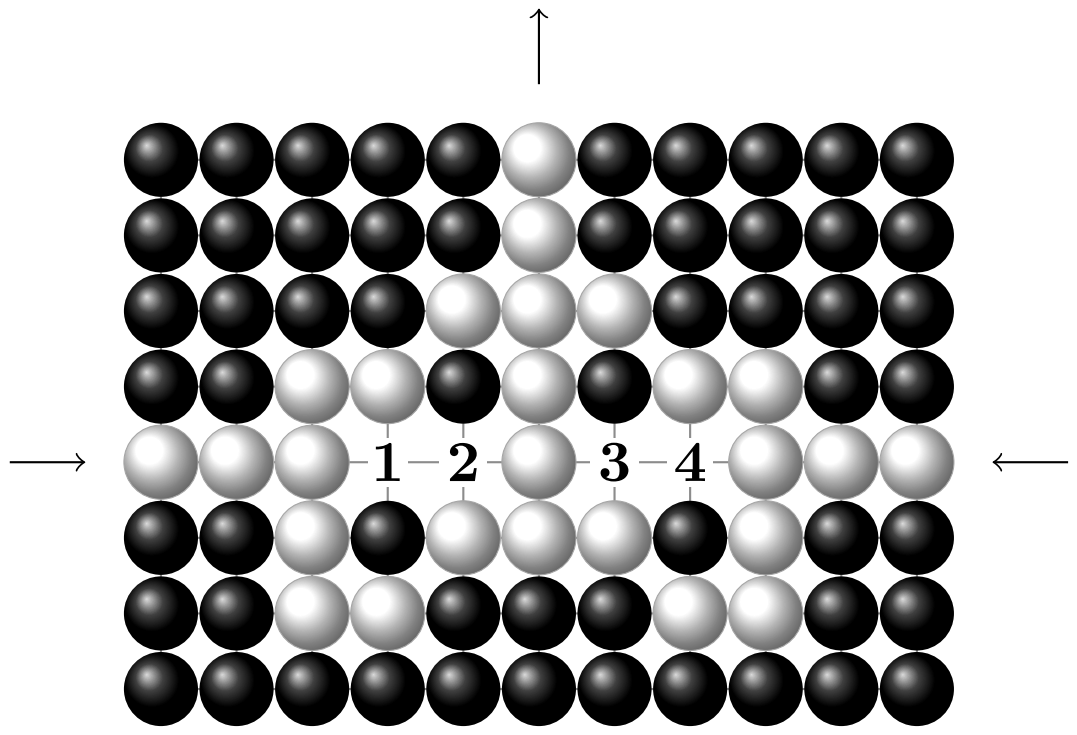

  \centering
  \GoBoardScaled{10}[7][.5]
    {(0,3), (1,3), (2,3), (2,2), (2,1), (3,1), (2,4), (3,4),
    (4,2), (5,2), (6,2), (5,3), (5,4), (5,5), (4,5), (6,5), (5,6), (5,7),
    (10,3), (9,3), (8,3), (8,4), (7,4), (8,2), (8,1), (7,1)}
    {(0,2), (1,2), (1,1), (1,0), (2,0), (3,0), (4,0), (4,1), (5,1), (6,1), (6,0), (7,0), (8,0), (9,0), (9,1), (9,2), (10,2),
    (3,2),
    (4,4),
    (7,2),
    (6,4),
    (10,4), (9,4), (9,5), (8,5), (7,5), (7,6), (6,6), (6,7),
    (4,7), (4,6), (3,6), (3,5), (2,5), (1,5), (1,4), (0,4),
    (0,0), (0,1),
    (5,0),
    (10,0), (10,1),
    (0, 5), (0,6), (0,7), (1,7), (1,6), (2,6), (2,7), (3,7),
    (7,7), (8,7), (8,6), (9,6), (9,7), (10,7), (10,6), (10,5)}
    [
      \draw[signal] (-2,3) -- (-1,3);
      \draw[signal] (12,3) -- (11,3);
      \draw[signal] (5,8) -- (5,9);
      \node[liberty] at (3,3) {1};
      \node[liberty] at (4,3) {2};
      \node[liberty] at (6,3) {3};
      \node[liberty] at (7,3) {4};
    ]
  \caption{OR gadget. The inputs are on the left and right sides; the output leaves via the top.}
  \label{fig:normalOr}
\end{figure}

In the OR gadget, shown in \cref{fig:normalOr}, there are two pairs of open
spots, labelled 1, 2, 3, and 4, all of which only Black can play in and only
once in each pair.  If both inputs are inactive, then Black can play in only 2
and 3, deactivating the output.  If either of the inputs (or both) is active,
then Black can play on at least one of 1 and 4, leaving the output active as
well.

\begin{figure}[htbp]
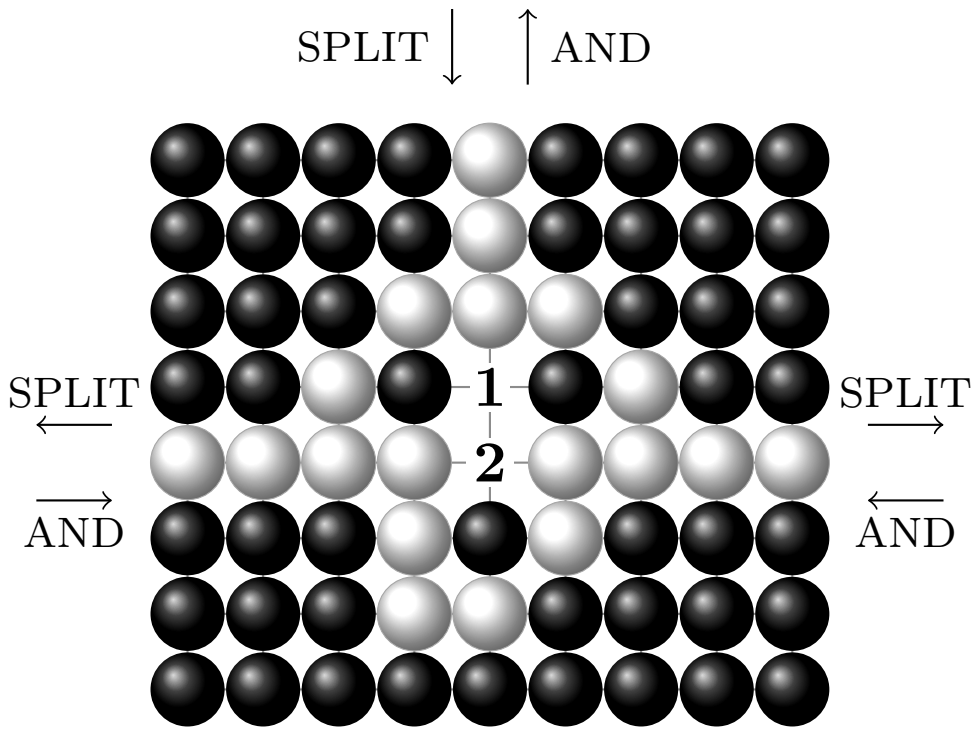

  \centering
  \GoBoardScaled{8}[7][.5]
    {(0,3), (1,3), (2,3), (2,4), (3,3), (3,2), (3,1), (4,1),
    (8,3), (7,3), (6,3), (6,4), (5,3), (5,2),
    (4,5), (3,5), (5,5), (4,6), (4,7)}
    {(0,2), (1,2), (2,2), (2,1), (2,0), (3,0), (4,0), (5,0), (5,1), (6,1), (6,2), (7,2), (8,2),
    (0,4), (1,4), (1,5), (2,5), (2,6), (3,6), (3,7),
    (5,7), (5,6), (6,6), (6,5), (7,5), (7,4), (8,4),
    (3,4), (5,4), (4,2),
    (0,0), (1,0), (1,1), (0, 1),
    (6,0), (7,0), (8,0), (8,1), (7,1),
    (0,5), (0,6), (0,7), (1,7), (1,6), (2,7),
    (6,7), (7,7), (7,6), (8,6), (8,5), (8,7)}
    [
      \draw[signal] (-1, 3.5) -- (-2, 3.5);
      \node[signal label, above=2pt] at (-1.5, 3.5) {SPLIT};
      \draw[signal] (-2, 2.5) -- (-1, 2.5);
      \node[signal label, below=2pt] at (-1.5, 2.5) {AND};

      \draw[signal] (9, 3.5) -- (10, 3.5);
      \node[signal label, above=2pt] at (9.5, 3.5) {SPLIT};
      \draw[signal] (10, 2.5) -- (9, 2.5);
      \node[signal label, below=2pt] at (9.5, 2.5) {AND};

      \draw[signal] (3.5, 9) -- (3.5, 8);
      \node[signal label, left=3pt] at (3.5, 8.5) {SPLIT};
      \draw[signal] (4.5, 8) -- (4.5, 9);
      \node[signal label, right=3pt] at (4.5, 8.5) {AND};

      \node[liberty] at (4,4) {1};
      \node[liberty] at (4,3) {2};
    ]
  \caption{
    Both the AND and SPLIT gadgets.  For AND, the inputs are on the left and
    right sides; the output leaves via the top.  For SPLIT, the input arrives
    from the top and the outputs leave on the sides.
  }
  \label{fig:normalAnd}
\end{figure}

The AND gadget, as shown in \cref{fig:normalAnd}, only has one open pair of spaces, labelled 1 and 2.  Only Black can play there and only in one of the two spaces.  If both inputs are active, then Black can play at 2 and activate the output.  Otherwise, if either input is inactive, then Black must play at 1, deactivating the output.

There is no need for a separate SPLIT gadget here, as we can reuse the AND gadget (\cref{fig:normalAnd}) with the inputs and outputs swapped.  If the input is active, then playing in 1 activates both outputs.  If the input is inactive, then Black has to play in 2, deactivating both outputs.

\begin{figure}[htbp]
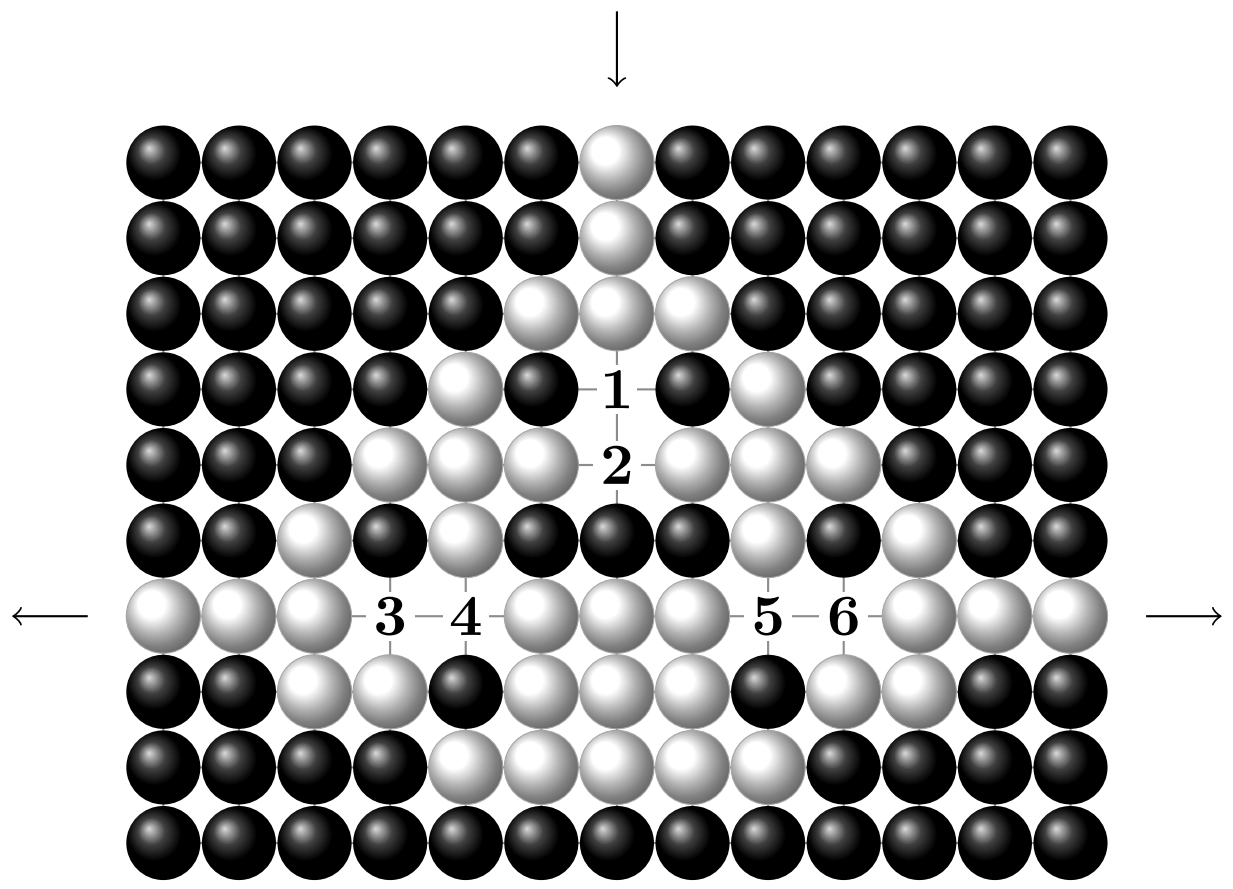

  \centering
  \GoBoardScaled{12}[9][.5]
    {(0,3), (1,3), (2,3), (2,4), (2,2), (3,2),
    (4,1), (5,1), (5,2), (5,3), (6,3), (7,3), (7,2), (6,2), (6,1), (7,1), (8,1),
    (9,2), (10,2), (10,3), (10,4), (11,3), (12,3),
    (3,5), (4,5), (4,4), (4,6), (5,5), (7,5), (8,5), (9,5), (8,4), (8,6),
    (5,7), (6,7), (7,7), (6,8), (6,9)}
    {(0,2), (1,2), (1,1), (2,1), (3,1), (3,0), (4,0), (5,0), (6,0), (7,0), (8,0), (9,0), (9,1), (10,1), (11,1), (11,2), (12,2),
    (4,2),
    (3,4),
    (5,6),
    (7,6),
    (9,4),
    (8,2),
    (5,4), (6,4), (7,4),
    (0,4), (1,4), (1,5), (2,5), (2,6), (3,6), (3,7), (4,7), (4,8), (5,8), (5,9),
    (7,9), (7,8), (8,8), (8,7), (9,7), (9,6), (10,6), (10,5), (11,5), (11,4), (12,4),
    (0,1), (0,0), (1,0), (2,0),
    (12,1), (12,0), (11,0), (10,0),
    (0,5), (0,6), (0,7), (0,8), (0,9), (1,9), (1,8), (1,7), (1,6), (2,7), (2,8), (2,9), (3,9), (3,8), (4,9),
    (8,9), (9,9), (9,8), (10, 7), (10,8), (10,9), (11,9), (11,8), (11,7), (11,6), (12,5), (12,6), (12,7), (12,8), (12,9)}
    [
      \draw[signal] (6, 11) -- (6, 10);
      \draw[signal] (-1, 3) -- (-2, 3);
      \draw[signal] (13, 3) -- (14, 3);
      \node[liberty] at (6,6) {1};
      \node[liberty] at (6,5) {2};
      \node[liberty] at (3,3) {3};
      \node[liberty] at (4,3) {4};
      \node[liberty] at (8,3) {5};
      \node[liberty] at (9,3) {6};
    ]
  \caption{CHOICE gadget.  The input enters through the top and the outputs exit via the sides.}
  \label{fig:normalChoice}
\end{figure}

In the CHOICE gadget, shown in \cref{fig:normalChoice}, there are three pairs
of open spaces, labelled 1 and 2, 3 and 4, and 5 and 6.  Only Black can play in
these pairs and only once in each.  If the input is inactive, then Black must
play in 2 and must then play in 3 and 6, deactivating the outputs.  If the
input is active, then Black can play in 1 and then either 4 and 6 (activating
the left output and deactivating the right) or 3 and 5 (activating the right
output and deactivating the left).

\begin{proof}[Proof of \cref{thm:square-grid} under normal play]
  \rsNoGo is in \PSPACE{} because on an $n \times n$ grid the game tree has
  depth and fanout at most $n^2$, and move legality can be checked in
  polynomial time, so we can traverse the entire game tree in \PSPACE.

  To show \PSPACE-hardness, we reduce from \rsBCL.  Using the gadgets described
  in this section so far, we can create a position where Black can play a
  number of times equal to
  \[
    x = a + 2b + c + d + 3e,
  \]
  where $a$ is the number of VARIABLE gadgets, $b$ is the number of
  OR gadgets, $c$ is the number of AND gadgets, $d$ is the number of SPLIT
  gadgets, and $e$ is the number of CHOICE gadgets. After play
  on the VARIABLE gadgets, Black will play first on the rest of the
  construction, so if we give White $x - a$ moves of their own,
  then Black will win if and only if they can get an extra move on the Goal
  gadget.  (White will have $a$ plays on the VARIABLE gadgets as well.) We can
  provide White with these moves by including a large block of white stones
  with $x + 1 - a$ singleton holes.  Only White can play in them and they can
  use all but one of them.  

  Thus, Black can win going first exactly when they can play the prescribed
  number of times on each gadget and also on the Goal gadget, which is exactly
  when they can win on the source \rsBCL position.

  \begin{figure}[htbp]
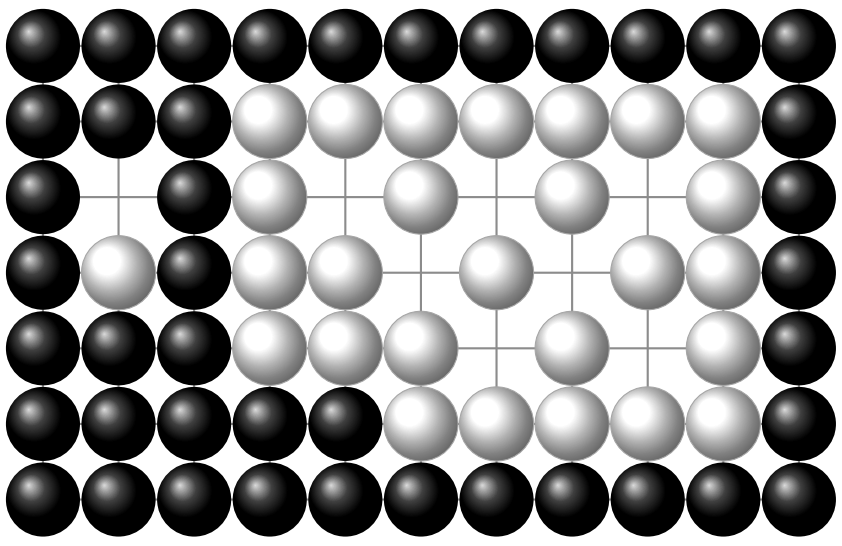

    \centering
    \GoBoardScaled{10}[6][.5]
      {(3,5), (4,5), (5,5), (6,5), (7,5), (8,5), (9,5),
       (3,4), (5,4), (7,4), (9,4),
       (3,3), (4,3), (6,3), (8,3), (9,3),
       (3,2), (4,2), (5,2), (7,2), (9,2),
       (5,1), (6,1), (7,1), (8,1), (9,1),
       (1,3)}
      {(0,6), (1,6), (2,6), (3,6), (4,6), (5,6), (6,6), (7,6), (8,6), (9,6), (10,6),
       (10,5), (10,4), (10,3), (10,2), (10,1), (10,0),
       (9,0), (8,0), (7,0), (6,0), (5,0), (4,0), (3,0), (2,0), (1,0), (0,0),
       (0,1), (0,2), (0,3), (0,4), (0,5),
       (1,5), (1,2), (1,1),
       (2,1), (2,2), (2,3), (2,4), (2,5),
       (3,1), (4,1)}
    \caption{
      A checkerboard white reserve inside the black background, together with a
      frozen liberty for the black background on the left. In this particular
      example, the white reserve replaces 33 black stones with 26 white stones
      and 7 interior holes, giving White 6 moves and Black none.
    }
    \label{fig:whiteMass}
  \end{figure}

  Because white wires can partition the black background into several regions,
  we give each black background component a frozen liberty, formed by a white
  singleton adjacent to a single empty point, all surrounded by black stones,
  as on the left of \cref{fig:whiteMass}; neither player can ever play there.
  For White's $x - a$ waiting moves, we place a checkerboard white reserve with
  $x + 1 - a$ pairwise non-adjacent interior holes surrounded by white stones,
  as on the right of \cref{fig:whiteMass} (a single inert interior hole when $x
  - a = 0$). Black can never play in these holes, while White can play in any
  $x - a$ of them before the last hole becomes illegal.

  To fit the construction onto an $n \times n$ grid with an equal number of
  black and white stones, we pad the board with black stones, choosing $n$
  large enough such that the number of occupied points is even and there is
  ample room outside of the gadgets. We can then easily thicken the white
  reserve with `inert' white stones (replacing adjacent black padding stones,
  without changing any empty points or disconnecting the black background from
  its frozen liberty) until the numbers of black and white stones are equal.
  Finally, any legal \rsNoGo position with an equal number of black and white
  stones is reachable from an empty $n \times n$ board with Black to move:
  deleting any stone preserves legality (every surviving component of its group
  gains the vacated point as a liberty), so by induction placing the stones
  alternately from the empty board stays legal at every step and leaves a
  position with Black to move. \end{proof}

\section{Mis\`ere play on the square grid}

In order to prove the mis\`ere-play version of \rsNoGo is \PSPACE-hard, we
will again reduce from \rsBCL, so we will need VARIABLE, AND, OR, SPLIT, and
CHOICE gadgets, as well as a Goal.

Just like with the normal-play hardness, the signals will be sent by wires of
white stones ``insulated'' by black stones with liberties at each end that
white cannot play at.  Unlike the normal-play version, the active signal sent
through the wires works oppositely: a wire is active if the input side is
blocked by a black stone, meaning the output side must be left open.

We reuse the Goal gadget from the normal-play construction (see
\cref{fig:normalGoal}).  Black will have an extra play if and only if the white
component connected to the input has a prior open liberty (if it is inactive).
(In the Goal gadget, neither player can play on the rightmost open space and
White can never play on the left open space.)

This means that if Black can activate the signal coming into the Goal, they
will be unable to play in the Goal.  As long as they have not cheated
elsewhere, not having that final play will mean that they will finish moving
before White and will win the game.

\begin{figure}[htbp]
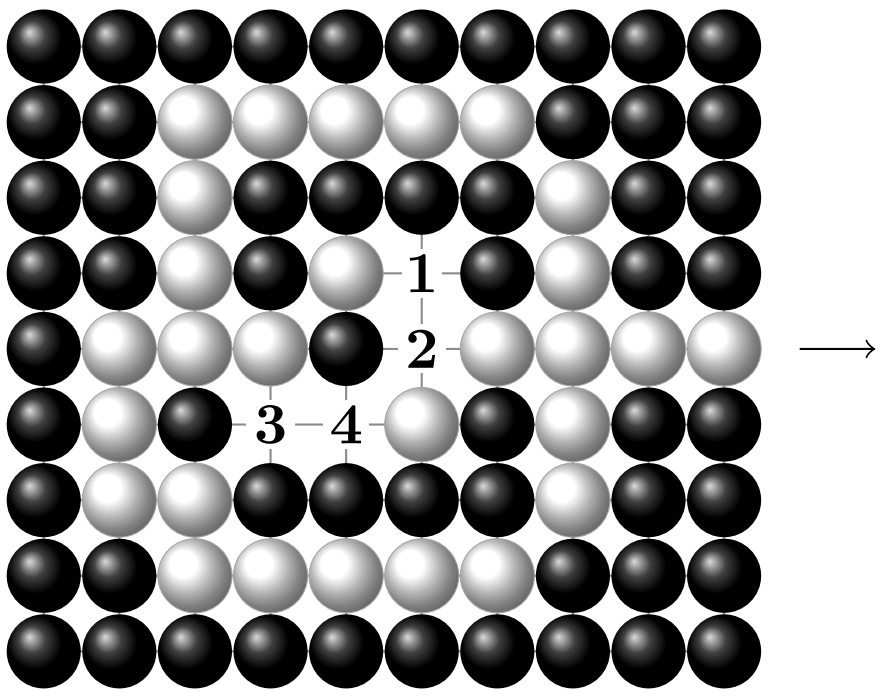

  \centering
  \GoBoardScaled{9}[8][.5]
    {(1,2), (1,3), (1,4),
    (2,1), (2,2), (2,4), (2,5), (2,6), (2,7),
    (3,1), (3,4), (3,7),
    (4,1), (4,5), (4,7),
    (5,1), (5,3), (5,7),
    (6,1), (6,4), (6,7),
    (7,2), (7,3), (7,4), (7,5), (7,6),
    (8,4),
    (9,4)}
    {(0,1), (0,2), (0,3), (0,4), (0,5),
    (1,0), (1,1), (1,5), (1,6), (1,7), (1,8),
    (2,0), (2,3), (2,8),
    (3,0), (3,2), (3,5), (3,6), (3,8),
    (4,0), (4,2), (4,4), (4,6), (4,8),
    (5,0), (5,2), (5,6), (5,8),
    (6,0), (6,2), (6,3), (6,5), (6,6), (6,8),
    (7,0), (7,1), (7,7), (7,8),
    (8,1), (8,2), (8,3), (8,5), (8,6), (8,7),
    (9,3), (9,5),
    (0,0),
    (8,0), (9,0), (9,1), (9,2),
    (0, 6), (0,7), (0,8),
    (9,8), (8,8), (9,7), (9,6)}
    [
      \draw[signal] (10,4) -- (11,4);
      \node[liberty] at (3,3) {3};
      \node[liberty] at (4,3) {4};
      \node[liberty] at (5,4) {2};
      \node[liberty] at (5,5) {1};
    ]
  \caption{VARIABLE gadget. The output leaves via the right-hand side.}
  \label{fig:misereVariable}
\end{figure}

The VARIABLE gadget, shown in \cref{fig:misereVariable}, has four open spaces (labelled 1, 2, 3, and 4) but neither player can play in spaces 1 or 3.  Of spaces 2 and 4, both players prefer to play in 2.  If White plays there, it leaves the output inactive.  If Black plays there, then the output is active.  Whichever player plays in 2, the other player will have a play in 4 which they can claim later in the game.

\begin{figure}[htbp]
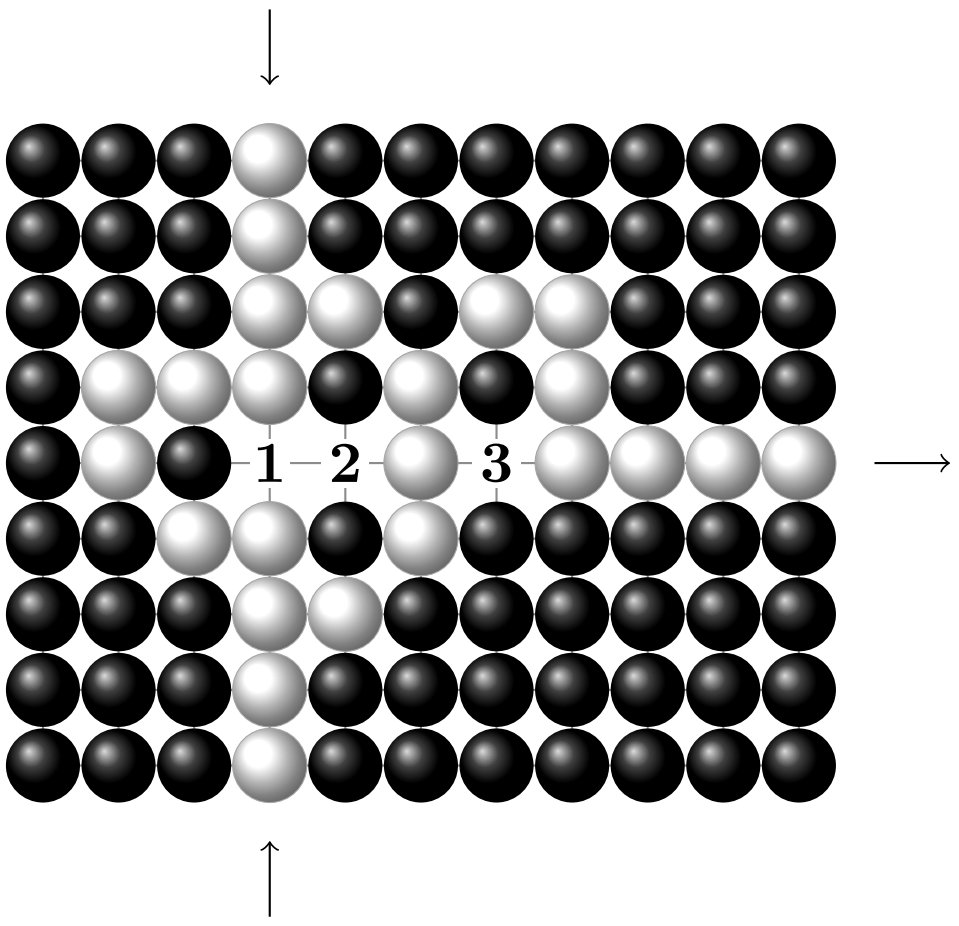

  \centering
  \GoBoardScaled{10}[8][.5]
    {(1,4), (1,5),
    (2,3), (2,5),
    (3,0), (3,1), (3,2), (3,3), (3,5), (3,6), (3,7), (3,8),
    (4,2), (4,6),
    (5,3), (5,4), (5,5),
    (6,6),
    (7,4), (7,5), (7,6),
    (8,4),
    (9,4),
    (10,4)}
    {(0,0), (0,1), (0,2), (0,3), (0,4), (0,5), (0,6), (0,7), (0,8),
    (1,0), (1,1), (1,2), (1,3), (1,6), (1,7), (1,8),
    (2,0), (2,1), (2,2), (2,4), (2,6), (2,7), (2,8),
    (4,0), (4,1), (4,3), (4,5), (4,7), (4,8),
    (5,0), (5,1), (5,2), (5,6), (5,7), (5,8),
    (6,0), (6,1), (6,2), (6,3), (6,5), (6,7), (6,8),
    (7,0), (7,1), (7,2), (7,3), (7,7), (7,8),
    (8,0), (8,1), (8,2), (8,3), (8,5), (8,6), (8,7), (8,8),
    (9,0), (9,1), (9,2), (9,3), (9,5), (9,6), (9,7), (9,8),
    (10,0), (10,1), (10,2), (10,3), (10,5), (10,6), (10,7), (10,8)}
    [
      \draw[signal] (3,10) -- (3,9);
      \draw[signal] (3,-2) -- (3,-1);
      \draw[signal] (11,4) -- (12,4);
      \node[liberty] at (3,4) {1};
      \node[liberty] at (4,4) {2};
      \node[liberty] at (6,4) {3};
    ]
  \caption{OR gadget.  The inputs come in from the top and bottom and the output leaves through the right.}
  \label{fig:misereOr}
\end{figure}

\cref{fig:misereOr} shows the OR gadget, which has three open spaces (labelled
1, 2, and 3) all only playable by Black. If either of the inputs is active,
then Black can only play in 2 or 3, but not both.  In that case, Black should
play in 3 to keep the output active.  If both of the inputs are inactive,
however, then Black can either legally play in both 1 and 3 or only 2.  Black
will choose the latter because it saves them a move, and the output will remain
inactive.

\begin{figure}[htbp]
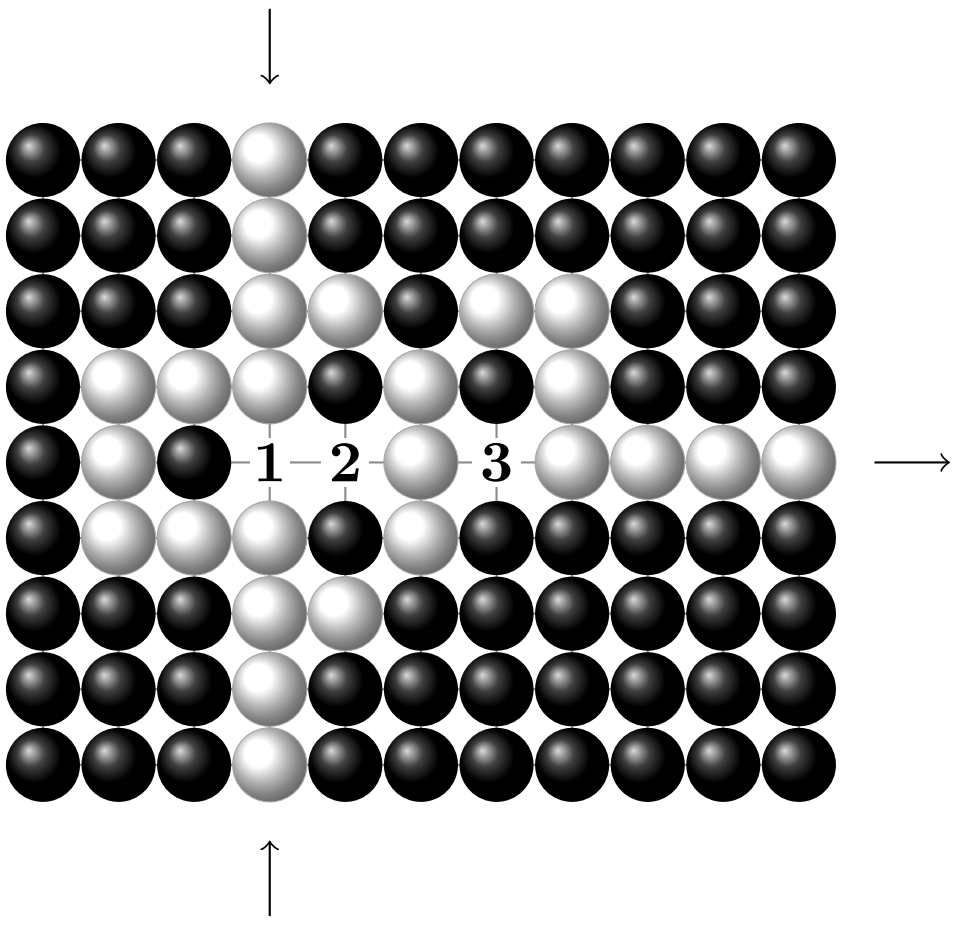

  \centering
  \GoBoardScaled{10}[8][.5]
    {(1,3), (1,4), (1,5),
    (2,3), (2,5),
    (3,0), (3,1), (3,2), (3,3), (3,5), (3,6), (3,7), (3,8),
    (4,2), (4,6),
    (5,3), (5,4), (5,5),
    (6,6),
    (7,4), (7,5), (7,6),
    (8,4),
    (9,4),
    (10,4)}
    {(0,0), (0,1), (0,2), (0,3), (0,4), (0,5), (0,6), (0,7), (0,8),
    (1,0), (1,1), (1,2), (1,6), (1,7), (1,8),
    (2,0), (2,1), (2,2), (2,4), (2,6), (2,7), (2,8),
    (4,0), (4,1), (4,3), (4,5), (4,7), (4,8),
    (5,0), (5,1), (5,2), (5,6), (5,7), (5,8),
    (6,0), (6,1), (6,2), (6,3), (6,5), (6,7), (6,8),
    (7,0), (7,1), (7,2), (7,3), (7,7), (7,8),
    (8,0), (8,1), (8,2), (8,3), (8,5), (8,6), (8,7), (8,8),
    (9,0), (9,1), (9,2), (9,3), (9,5), (9,6), (9,7), (9,8),
    (10,0), (10,1), (10,2), (10,3), (10,5), (10,6), (10,7), (10,8)}
    [
      \draw[signal] (3,10) -- (3,9);
      \draw[signal] (3,-2) -- (3,-1);
      \draw[signal] (11,4) -- (12,4);
      \node[liberty] at (3,4) {1};
      \node[liberty] at (4,4) {2};
      \node[liberty] at (6,4) {3};
    ]
  \caption{AND gadget. The inputs come from the top and bottom and the output leaves on the right.}
  \label{fig:misereAnd}
\end{figure}

The AND gadget (shown in \cref{fig:misereAnd}) is very similar to the OR,
except with one stone switched from black to white.  This gadget also has three
open spaces (labelled 1, 2, and 3) only playable by Black.  This time if either
of the inputs is inactive, then 1 is playable (so playing 3 would leave an
extra move at 1), so Black should play on space 2 and leave the output
inactive.  If both inputs are active, then Black can play on 3, leaving 1 and 2
unplayable and activating the output.

\begin{figure}[htbp]
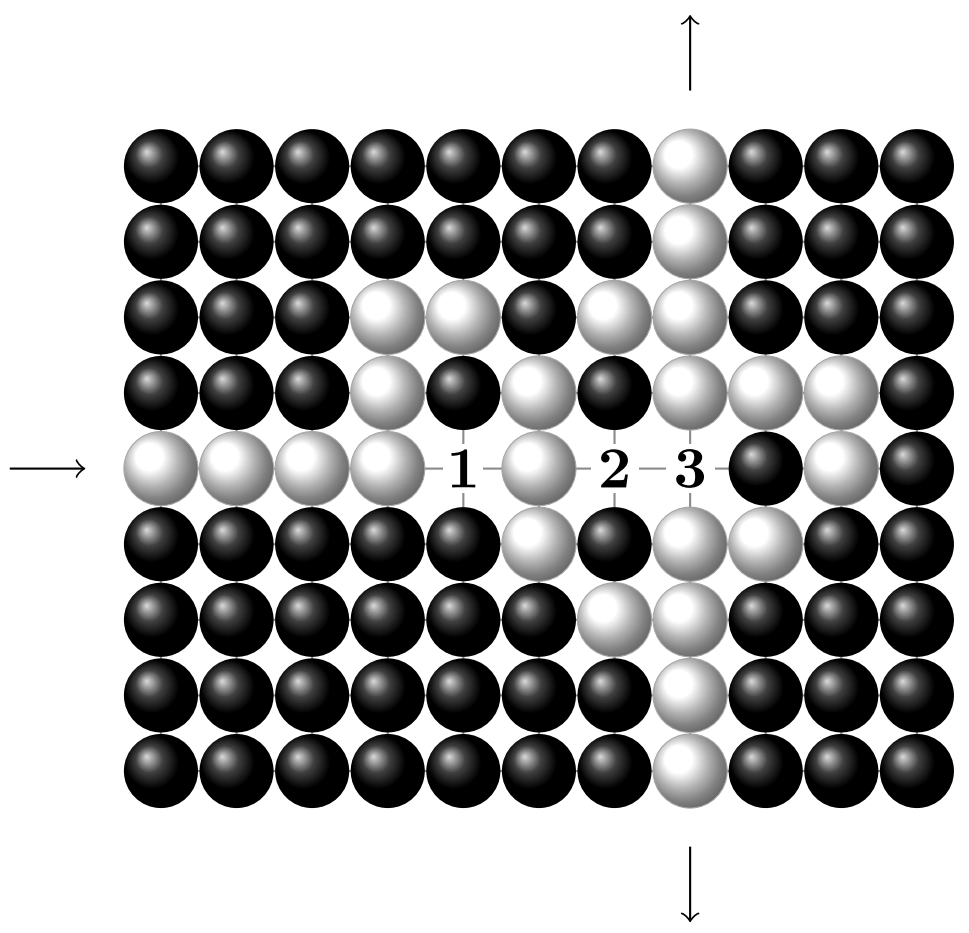

  \centering
  \GoBoardScaled{10}[8][.5]
    {(0,4),
    (1,4),
    (2,4),
    (3,4), (3,5), (3,6),
    (4,6),
    (5,3), (5,4), (5,5),
    (6,2), (6,6),
    (7,0), (7,1), (7,2), (7,3), (7,5), (7,6), (7,7), (7,8),
    (8,3), (8,5),
    (9,4), (9,5)}
    {(0,0), (0,1), (0,2), (0,3), (0,5), (0,6), (0,7), (0,8),
    (1,0), (1,1), (1,2), (1,3), (1,5), (1,6), (1,7), (1,8),
    (2,0), (2,1), (2,2), (2,3), (2,5), (2,6), (2,7), (2,8),
    (3,0), (3,1), (3,2), (3,3), (3,7), (3,8),
    (4,0), (4,1), (4,2), (4,3), (4,5), (4,7), (4,8),
    (5,0), (5,1), (5,2), (5,6), (5,7), (5,8),
    (6,0), (6,1), (6,3), (6,5), (6,7), (6,8),
    (8,0), (8,1), (8,2), (8,4), (8,6), (8,7), (8,8),
    (9,0), (9,1), (9,2), (9,3), (9,6), (9,7), (9,8),
    (10,0), (10,1), (10,2), (10,3), (10,4), (10,5), (10,6), (10,7), (10,8)}
    [
      \draw[signal] (-2,4) -- (-1,4);
      \draw[signal] (7,9) -- (7,10);
      \draw[signal] (7,-1) -- (7,-2);
      \node[liberty] at (4,4) {1};
      \node[liberty] at (6,4) {2};
      \node[liberty] at (7,4) {3};
    ]
  \caption{SPLIT gadget.  The input arrives from the left and the outputs leave via the top and bottom.}
  \label{fig:misereSplit}
\end{figure}

The SPLIT gadget is shown in \cref{fig:misereSplit}, with three open spaces
playable by Black only, labelled 1, 2, and 3.  If the input is active, then
Black only has one play in the gadget, either at 2 or 3.  Black should choose
to play on 3 to activate both outputs.  If the input is inactive, then all
three spaces are playable, so Black should choose to play on 2 (instead of both
1 and 3), which leaves the outputs inactive.

\begin{figure}[htbp]
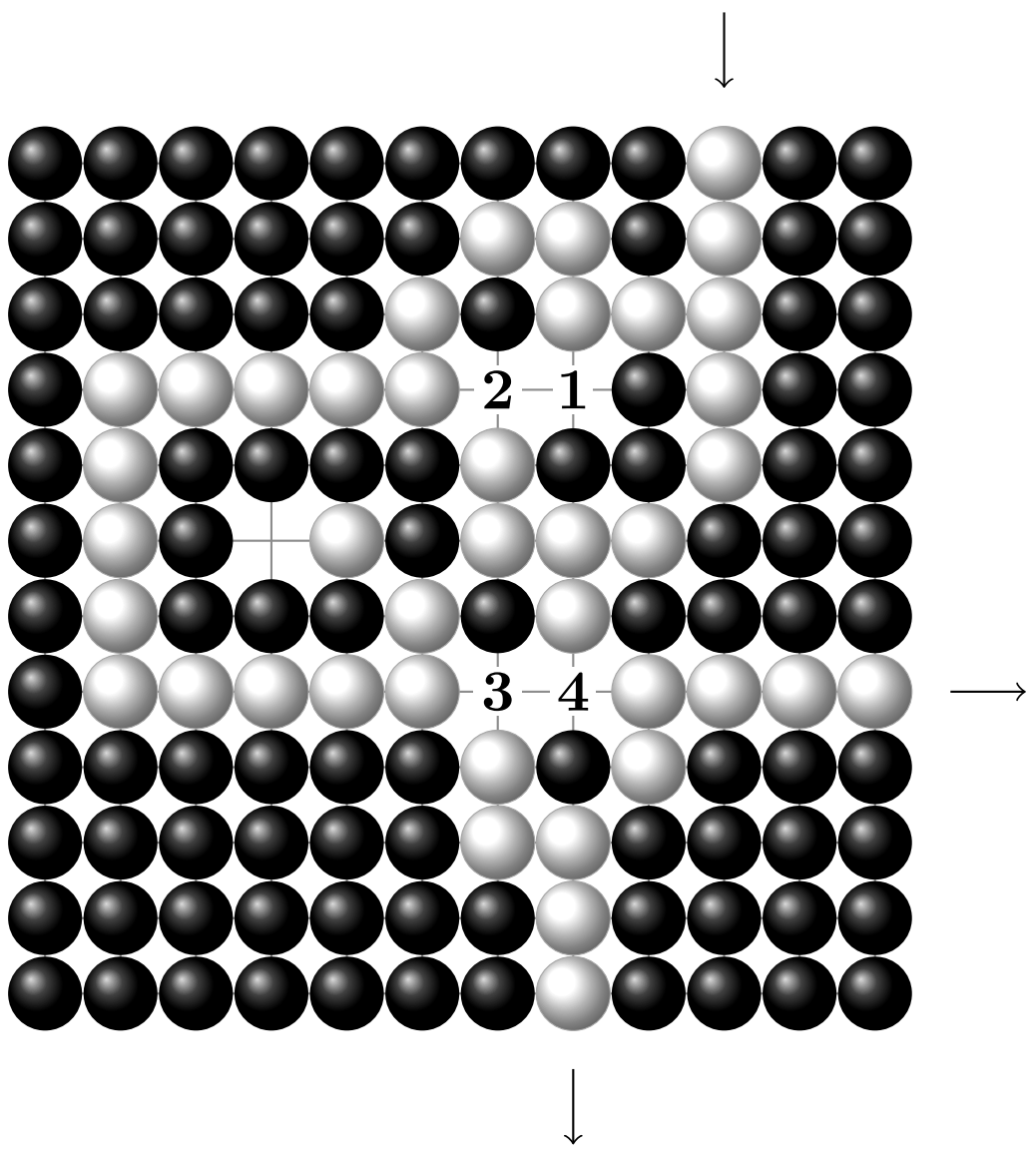

  \centering
  \GoBoardScaled{11}[11][.5]
    {(1,4), (1,5), (1,6), (1,7), (1,8),
    (2,4), (2,8),
    (3,4), (3,8),
    (4,4), (4,6), (4,8),
    (5,4), (5,5), (5,8), (5,9),
    (6,2), (6,3), (6,6), (6,7), (6,10),
    (7,0), (7,1), (7,2), (7,5), (7,6), (7,9), (7,10),
    (8,3), (8,4), (8,6), (8,9),
    (9,4), (9,7), (9,8), (9,9), (9,10), (9,11),
    (10,4),
    (11,4)}
    {(0,0), (0,1), (0,2), (0,3), (0,4), (0,5), (0,6), (0,7), (0,8), (0,9), (0,10), (0,11),
    (1,0), (1,1), (1,2), (1,3), (1,9), (1,10), (1,11),
    (2,0), (2,1), (2,2), (2,3), (2,5), (2,6), (2,7), (2,9), (2,10), (2,11),
    (3,0), (3,1), (3,2), (3,3), (3,5), (3,7), (3,9), (3,10), (3,11),
    (4,0), (4,1), (4,2), (4,3), (4,5), (4,7), (4,9), (4,10), (4,11),
    (5,0), (5,1), (5,2), (5,3), (5,6), (5,7), (5,10), (5,11),
    (6,0), (6,1), (6,5), (6,9), (6,11),
    (7,3), (7,7), (7,11),
    (8,0), (8,1), (8,2), (8,5), (8,7), (8,8), (8,10), (8,11),
    (9,0), (9,1), (9,2), (9,3), (9,5), (9,6),
    (10,0), (10,1), (10,2), (10,3), (10,5), (10,6), (10,7), (10,8), (10,9), (10,10), (10,11),
    (11,0), (11,1), (11,2), (11,3), (11,5), (11,6), (11,7), (11,8), (11,9), (11,10), (11,11)}
    [
      \draw[signal] (9,13) -- (9,12);
      \draw[signal] (7,-1) -- (7,-2);
      \draw[signal] (12,4) -- (13,4);
      \node[liberty] at (7,8) {1};
      \node[liberty] at (6,8) {2};
      \node[liberty] at (6,4) {3};
      \node[liberty] at (7,4) {4};
    ]
  \caption{CHOICE gadget.  The input arrives from the top and the outputs leave via the right and bottom.}
  \label{fig:misereChoice}
\end{figure}

The CHOICE gadget, shown in \cref{fig:misereChoice}, has five open spaces, but one cannot be played by either player.  The remaining four (labelled 1, 2, 3, and 4) are all playable only by Black.  If the input is active, then 1 is unplayable and Black can choose to play at either 3 or 4, activating the corresponding output and leaving 2 unplayable.  If the input is inactive, then if Black plays at either 3 or 4, 1 will still be playable and Black will need to make an extra move.  If they play at 2, however, then both 3 and 4 are unplayable and both outputs are inactive.

\begin{proof}[Proof of \cref{thm:square-grid} under mis\`ere play]
  \rsNoGo is in \PSPACE{} because on an $n \times n$ grid the game tree has
  depth and fanout at most $n^2$, and move legality can be checked in
  polynomial time, so we can traverse the entire game tree in \PSPACE, just as
  with the normal-play version.

  To show \PSPACE-hardness, we reduce from \rsBCL. Using the gadgets described
  earlier in this section, we construct a position where Black's intended
  number of moves across the non-Goal gadgets is
  \[
    x = a + b + c + d + e,
  \]
  where $a, b, c, d, e$ are the numbers of VARIABLE, OR, AND, SPLIT, and CHOICE
  gadgets, respectively. We give White $x - a$ additional moves using a
  checkerboard white reserve with $x + 1 - a$ interior holes (as in
  \cref{fig:whiteMass}), so White has $x$ moves in total ($a$ on the VARIABLE
  gadgets and $x - a$ in the reserve). Since Black moves first, Black wins
  under mis\`ere play if and only if Black runs out of moves after at most
  $x$ plays. In each logic gate, space 2 remains legal until Black plays in
  that gate (forcing at least one Black move per gate), and White can likewise
  force at least one Black move in each VARIABLE gadget by playing 2 on
  Black-unused VARIABLEs and playing 4 only after Black plays 2. Thus Black can
  exhaust in at most $x$ moves precisely when Black can spend one move per
  gadget and activate the signal into Goal (avoiding an extra move there),
  which occurs if and only if Black wins the source \rsBCL position.

  To make the reduced positions reachable from an empty $n \times n$ grid with
  Black to move, we pad the board and thicken the white reserve in the same way
  as in the proof for normal play.
\end{proof}

\section{Subcubic hardness}

We prove \cref{thm:subcubic} by reducing from \rsCAL. Recall that \rsCAL is an
impartial ruleset played on a multigraph: vertices are called \emph{coins} and
the edges are called \emph{strings}. On a player's turn, they must cut a string
(i.e.\ delete an edge). If a player removes the last string attached to a coin,
thus freeing the coin, they lose. Equivalently, we can forbid moves that free a
coin to use the normal play perspective: if a player cannot make a legal move,
then they lose. At the end of this section, we will prove the following lemma;
we defer its proof for now so that we can first show the reduction for
\rsGNoGo.

\begin{lemma}
  \label{lem:coins-are-lava}
  \rsCAL is \PSPACE-complete under normal play on connected bipartite
  multigraphs with no isolated vertices, given as lists of their strings.
\end{lemma}

\cref{lem:coins-are-lava} is all we need to prove subcubic hardness for
\rsGNoGo. Keep in mind that our overall strategy for this reduction is to
replace coins with groups and strings with liberties; we use the bipartiteness
guaranteed by \cref{lem:coins-are-lava} to ensure that strings represent
liberties shared by opposite-coloured groups.

\begin{proof}[Proof of \cref{thm:subcubic}]
  Let $G$ be a connected bipartite multigraph (with no isolated vertices).
  Colour the coins in one partite set Black, and the other partite set White.
  Replace each coin of degree $d$ by a path of $2d-1$ stones of its colour, and
  assign its $d$ strings to every other stone (first, third, et cetera), up
  until the last stone of the path. Then, for each string, add one empty point
  adjacent to the two stones assigned to it. See \cref{fig:subcubic} for an
  explicit example.

  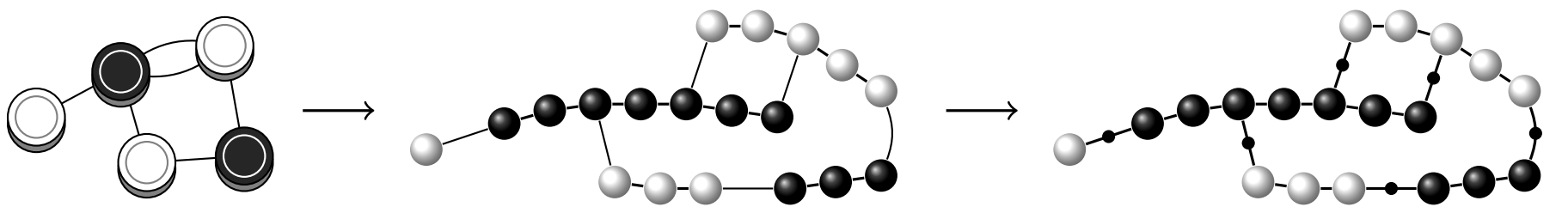
\begin{figure}[htb]
    \centering
    \begin{tikzpicture}[
      x=1mm, y=1mm,
      string/.style={thin},
      edge/.style={semithick},
      stone/.style={circle, inner sep=0pt, minimum size=2.5mm},
      black stone/.style={stone, ball color=black},
      white stone/.style={stone, ball color=white},
      vacant/.style={circle, fill, inner sep=0pt, minimum size=1mm}
      ]
      \coordinate (a) at (6.5,9);
      \coordinate (b) at (14.5,11);
      \coordinate (c) at (16,2.5);
      \coordinate (d) at (8.5,2);
      \coordinate (e) at (0,5.5);
      \draw[string] (e) -- (a)  (a) -- (d)  (b) -- (c)  (c) -- (d)
        (a) to[bend left] (b)  (a) to[bend right] (b);

      \foreach \coin in {a, c} {
        \draw[fill=gray] (\coin) ++(0,-.5) circle[radius=2.2mm];
        \draw[fill=black!85] (\coin) circle[radius=2.2mm];
        \draw[white, thin] (\coin) circle[radius=1.6mm];
      }
      \foreach \coin in {b, d, e} {
        \draw[fill=gray] (\coin) ++(0,-.5) circle[radius=2.2mm];
        \draw[fill=white] (\coin) circle[radius=2.2mm];
        \draw[gray, thin] (\coin) circle[radius=1.6mm];
      }

      \newcommand\stones{
        \node[white stone] (e1) at (0,3) {};
        \path[nodes=black stone]
          (6,5) node (a1) {}  (9.5,6) node (a2) {}  (13,6.5) node (a3) {}
          (16.5,6.5) node (a4) {}  (20,6.5) node (a5) {}
          (23.5,6) node (a6) {}  (27,5.5) node (a7) {};
        \path[nodes=white stone]
          (22,12.5) node (b1) {}  (25.5,12.5) node (b2) {}
          (29,11.5) node (b3) {}  (32,9.5) node (b4) {}  (35,7.5) node (b5) {};
        \path[nodes=black stone]
          (28,0) node (c1) {}  (31.5,.5) node (c2) {}  (35,1) node (c3) {};
        \path[nodes=white stone]
          (14.5,.5) node (d1) {}  (18,0) node (d2) {}  (21.5,0) node (d3) {};
        \draw[edge] (a1) -- (a2) -- (a3) -- (a4) -- (a5) -- (a6) -- (a7)
          (b1) -- (b2) -- (b3) -- (b4) -- (b5)  (c1) -- (c2) -- (c3)
          (d1) -- (d2) -- (d3);
      }

      \begin{scope}[xshift=30mm]
        \draw[->, semithick] (-9.5,6) -- (-4,6);
        \stones
        \draw[string] (e1) -- (a1);
        \draw[string] (a3) -- (d1);
        \draw[string] (a5) -- (b1);
        \draw[string] (a7) -- (b3);
        \draw[string] (b5) to[bend left=20] (c3);
        \draw[string] (c1) -- (d3);
      \end{scope}

      \begin{scope}[xshift=79.5mm]
        \draw[->, semithick] (-9.5,6) -- (-4,6);
        \stones
        \draw[edge] (e1) -- node[vacant] {} (a1);
        \draw[edge] (a3) -- node[vacant] {} (d1);
        \draw[edge] (a5) -- node[vacant] {} (b1);
        \draw[edge] (a7) -- node[vacant] {} (b3);
        \draw[edge] (b5) to[bend left=20] node[vacant] {} (c3);
        \draw[edge] (c1) -- node[vacant] {} (d3);
      \end{scope}
    \end{tikzpicture}
    \caption{
      From bipartite \rsCAL to bipartite \rsGNoGo. We colour the coins black
      and white only to illustrate the bipartition.
    }
    \label{fig:subcubic}
  \end{figure}

  The resulting board is necessarily connected and simple, even if two coins of
  $G$ are joined by several strings. Furthermore, each stone has at most two
  neighbours along its path, and at most one empty neighbour corresponding to a
  string, while each empty point has degree 2; as such, the board is subcubic.
  Finally, we can properly 2-colour the graph by putting in one partite set all
  of the stones that had a string assigned to them, and the rest into the other
  partite set; thus, the graph is bipartite.

  Initially, each coin in our original multigraph is represented by a single
  group in our resulting \rsGNoGo position, and the liberties of the group are
  precisely the strings attached to the coin. Note that no coin is isolated,
  and so the position is a legal \rsGNoGo position. We will think of a string
  as having been cut if its associated liberty has been filled (by either
  player). We claim that, throughout play, the liberties of each group
  correspond precisely to the uncut strings of its associated coin.

  To see this, consider the point representing a string between two coins. It
  is adjacent to one stone of each colour. A stone played on the liberty (by
  either player) joins the single adjacent group of its own colour and takes
  one liberty away from each of the two groups. It adds no new liberties, since
  neither neighbour is empty. Thus, either player may play on the liberty
  precisely when both groups already have another liberty.

  Hence, Black and White have the same move set after every run of play: the
  game tree is impartial. And the uncut strings determine the moves in the way
  we have described. The construction takes linear time in the size of the
  given multigraph, so \cref{lem:coins-are-lava} yields \PSPACE-hardness.
  Membership in \PSPACE{} is immediate, since each move fills a point (and we
  can check for legal moves by finding the groups and their liberties). Thus,
  we have the result.
\end{proof}

Since a game is equal to 0 precisely when the previous player wins, one can
observe that \cref{thm:subcubic} also tells us that it is \PSPACE-complete to
decide whether one of those impartial positions (satisfying the hypotheses) is
0. (This follows from \PSPACE{} being closed under complementation.)

For $\ell \ge 1$, the \emph{uniform $\ell$-subdivision} of a graph replaces
every edge by a path of $\ell$ internal vertices (of degree 2).

\begin{corollary}
  \label{cor:subdivisions}
  For a fixed $\ell\ge1$, \rsGNoGo remains \PSPACE-complete under normal play
  on uniform $\ell$-subdivisions of connected bipartite subcubic graphs, even
  for positions whose game trees are impartial.
\end{corollary}

\begin{proof}
  Consider an edge of the board constructed in the proof of
  \cref{thm:subcubic}. It is incident with at least one stone; and, if both of
  its endpoints are stones, then they have the same colour by construction. Now
  consider the $\ell$-subdivision of this edge and fill all of the new points
  with the same colour that the edge was incident to previously. Following this
  procedure on every edge extends each group to the same liberties as before,
  and so the entire game tree is preserved. This construction only takes
  polynomial time, and hence we have the result.
\end{proof}

Together, \cref{thm:subcubic,cor:subdivisions} give conditions C2 and C3 of
Johnson et al.~\cite{johnson.ea:C123}. We spell out the resulting
forbidden-subgraph consequence that is immediate from their framework and their
results. Note that a subdivided claw is obtained from $K_{1,3}$ by subdividing
each edge an arbitrary number of times (i.e.\ it is a three-legged spider). We
say that a graph is $\mathcal{H}$-subgraph-free if it contains no member of
$\mathcal{H}$ as a subgraph.

\begin{corollary}
  \label{cor:forbidden}
  If $\mathcal{H}$ is a fixed finite family of non-empty graphs, none of which
  is a disjoint union of paths and subdivided claws, then \rsGNoGo is
  \PSPACE-complete under normal play on $\mathcal{H}$-subgraph-free graphs.
\end{corollary}

\begin{proof}
  Choose $\ell\ge1$ to be larger than the number of vertices of every graph in
  $\mathcal{H}$. On the boards in the hypothesis of \cref{cor:subdivisions},
  every cycle has more than $\ell$ vertices, and any two vertices of degree
  3 are at distance greater than $\ell$. Hence, every connected subgraph on at
  most $\ell$ vertices is a subcubic tree with at most one branching vertex.
  Such a tree is necessarily a path or a subdivided claw. Thus, no member of
  $\mathcal{H}$ is a subgraph of these boards, and thus we have the result by
  \cref{cor:subdivisions}.
\end{proof}

\subsection{\rsCAL on bipartite multigraphs}

To prove \cref{lem:coins-are-lava}, we reduce from Schaefer's positive-CNF
game~\cite[Theorem~3.6]{DBLP:journals/jcss/Schaefer78}. Black and White take
turns claiming variables, with Black setting each claimed variable to True and
White setting each claimed variable to False. Black wins if the resulting
assignment satisfies the given formula.

Before we prove the result in full, we will first walk through a simple example
to illustrate the construction; this will make understanding the general
construction more clear. Consider the formula $(x\lor y)\land z$. Going first,
Black can secure a win by first claiming $z$; after White claims one of $x$ or
$y$, Black will claim the other one, resulting in a True formula.

To represent this formula game in \rsCAL, we write down three rows:
\begin{align*}
  A&=(x,y,1),\\
  B&=(z),\\
  C&=(0).
\end{align*}
We call a row True if a majority of its entries are True. The extra 1 in row
$A$ then means that the row is True (i.e.\ has majority) if and only if at
least one of $x$ or $y$ is True; that is, if and only if $x\lor y$. Row $B$ has
the logical value of $z$. Since row $C$ is always False, a majority of the rows
are True if and only if both $A$ and $B$ are True, which occurs if and only if
$(x\lor y)\land z$, which is our formula that we are attempting to emulate.

\begin{figure}[htb]
  \centering
  \begin{tikzpicture}[
    x=7mm, y=7mm, label distance=2pt,
    rope/.style={double, thick},
    width/.style={
      fill=white, rounded corners=4pt, inner sep=1.5pt, font=\footnotesize
    }
  ]
    % root and rows
    \coordinate[label=above:$r$] (r) at (8,6);
    \foreach \row/\col/\mid/\side in
      {B/0/4/left, A/8/8/above right, C/14/11/right} {
      \coordinate[label=\side:$\row$] (\row) at (\col,2);
      \coordinate (mid-\row) at (\mid,4);
      \draw[rope] (r) -- node[width] {$N$} (mid-\row);
      \draw[rope] (mid-\row) -- node[width] {$N^2$} (\row);
    }

    % variable paths
    \foreach \var/\col in {z/0, x/4, y/8} {
      \coordinate[label=left:$\var$] (\var) at (\col,0);
      \coordinate (mid-\var) at ({\col+1},0);
      \coordinate (end-\var) at ({\col+2},0);
      \draw (\var) -- (mid-\var) -- (end-\var);
    }

    % constant path
    \coordinate[label=right:$1$] (one) at (12,0);
    \coordinate (mid-const) at (13,0);
    \coordinate[label=right:$0$] (zero) at (14,0);
    \draw (mid-const) -- (zero);

    % row entries
    \foreach \row/\entry in {B/z, A/x, A/y, A/one, C/zero}
      \draw[rope] (\row) -- node[width] {$N^3$} (\entry);

    % path ends and leaf
    \coordinate[label=below right:$g$] (g) at (7,-4);
    \coordinate (leaf) at (7,-5);
    \draw (g) -- (leaf);
    \foreach \tip in {z, end-z, x, end-x, y, end-y, one, zero}
      \draw[rope] (g) -- node[width] {$N^4$} (\tip);

    % coins
    \foreach \coin in {r, A, B, C, mid-A, mid-B, mid-C,
        z, mid-z, end-z, x, mid-x, end-x, y, mid-y, end-y,
        one, mid-const, zero, g, leaf}
      \fill (\coin) circle[radius=2pt];
  \end{tikzpicture}
  \caption{
    The construction for $(x\lor y)\land z$ in the reduction in the proof of
    \cref{lem:coins-are-lava}. A single line is a single string, whereas a
    double line labelled $w$ is a rope of width $w$.
  }
  \label{fig:cal-example}
\end{figure}
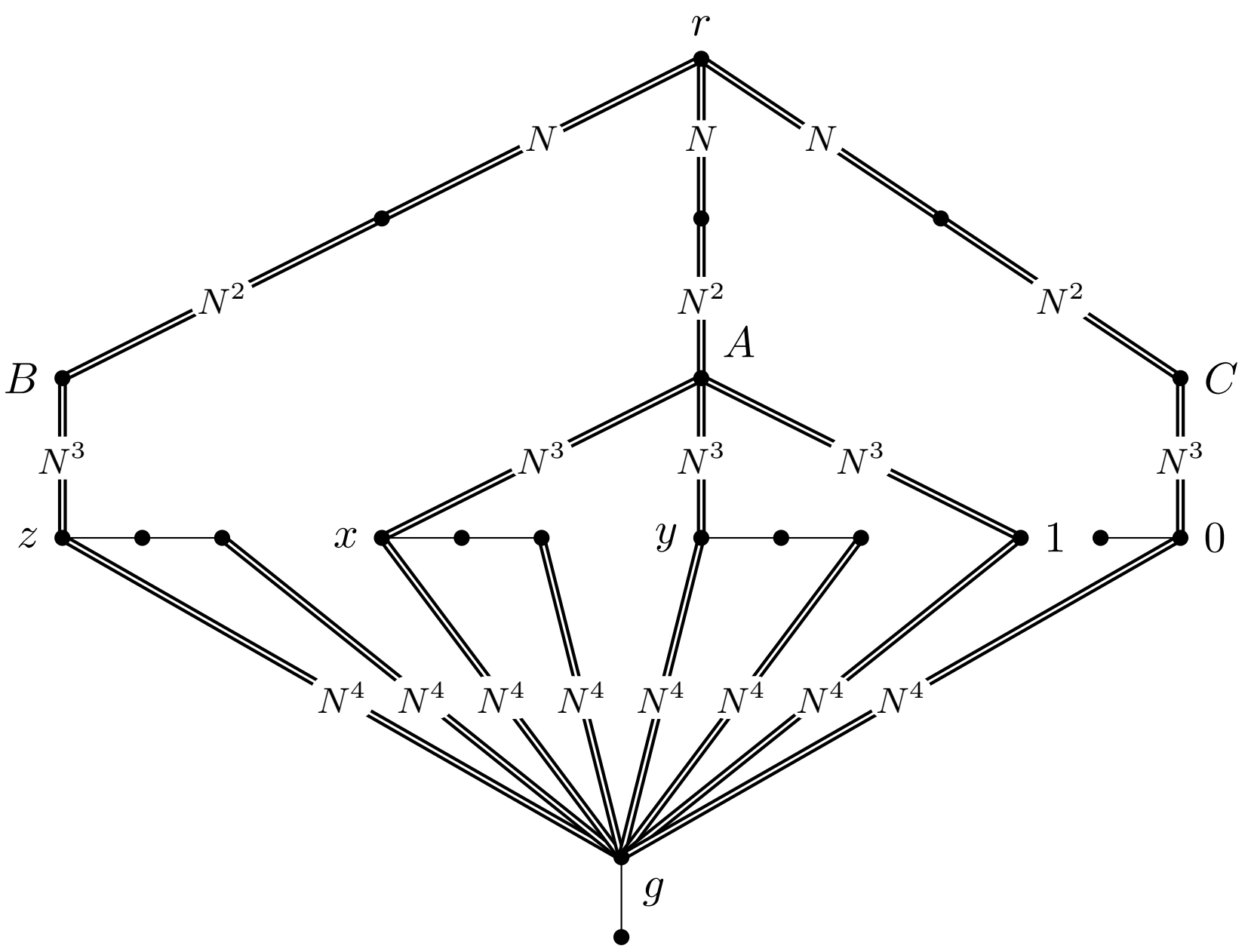

The \rsCAL position that we construct for this formula is shown in
\cref{fig:cal-example}. Each double line represents a \emph{rope} of
potentially many strings (i.e.\ a multiedge), and its \emph{width} is the
number of strings it contains. Each row is joined to the root $r$ by a
\emph{wire}~\cite[Figure~5]{demaine.diomidova:strings-and-coins}, which
consists of two ropes meeting at a middle coin; the rope on the $r$ side has
width $N$, and the rope on the row side has width $N^2$. From each row vertex,
a rope of width $N^3$ then leads to each of its entries. For instance, the
three ropes from $A$ lead to all of $x$, $y$, and 1.

The procedure to keep in mind is as follows: the player who can win the formula
game will first play on the variable paths according to their winning strategy.
This gives Black a majority of True rows if Black can win the formula game, or
White a majority of False rows if White can win. The winning player then uses
this majority at the root: they cut towards the rows of the wrong value (i.e.\
Black cuts wires that lead to False, and White cuts wires that lead to True),
which forces the last string remaining at $r$ to lead towards a row of their
desired value. They then use the same idea again at the row: for example, if
Black wins, then Black will force a True row, which itself has a majority of
True entries, and so Black can play to the majority again, cutting against the
False values; in this way, Black would leave the last string at the row to a
True value. Now, the parity is set up so that a True entry gives an odd number
of cuts and a False entry gives an even number, and so in either case the
winner of the formula game makes the last cut. Let us see how Black does this
on our $(x\lor y)\land z$ example.

Each variable has a path of two single strings (a $P_3$). Cutting the string at
the labelled end sets the variable to True, while cutting the other sets it to
False; note that the middle coin of the path is only adjacent to these two
strings, so they cannot both be cut. The constants 1 and 0 are drawn in a
special form: 1 can be considered as a variable with its True edge already cut,
and 0 is a variable where the False edge is already cut, but some of the extra
bits can be omitted, as shown in the construction.

Both ends of every variable path are joined to the `ground' vertex $g$ by ropes
of width $N^4$, and $g$ has an extra leaf of its own (which can never be cut,
since it would isolate said leaf).

Now, suppose that Black cuts the string at $z$, and White cuts the string at
the unlabelled (False) end of the $y$-path, and then Black cuts the string at
$x$. Then $x$ and $z$ are True, $y$ is False, and the two True rows are $A$ and
$B$. Note that we do not actually require White to play along with us on these
variables; indeed, they can play wherever they like, and Black could just spend
an extra turn making another variable True. The point is that we make the ropes
thick enough so that it does not matter where White plays; more generally, the
winner of the positive-CNF formula will always follow the strategy we describe,
regardless of what their opponent does (and, in the case of our worked example
$(x\lor y)\land z$ here, the winner is Black).

After claiming variables in the way we described (so that they win the
formula), Black turns their attention to the ropes of width $N$; their aim now
is to force the last string at $r$ to lead towards a True row (i.e.\ one of $A$
or $B$). Initially, $2N$ strings at $r$ lead to $A$ or $B$, whereas only $N$
lead towards $C$. Black will therefore cut towards $C$ on every move until it
is gone. Regardless of how White plays, at least one string pointing towards a
True row remains. Black then continues to cut arbitrarily at $r$ until just one
string remains. Suppose that the final string leads towards $A$.

Black will now cut right through the rope of width $N^2$ between $A$ and the
middle coin of its wire. The last string at $r$ holds the middle coin, and the
ropes to $x$, $y$, and 1 also hold $A$, so Black can cut all such strings
legally. After this, Black uses the same majority procedure again: they want to
ensure that the last string adjacent to $A$ is connected to a True part of the
clause (i.e.\ to $x$ or 1 in our example here). Note that $x$ and 1 started
with $2N^3$ strings in total, against just $N^3$ for the False entry $y$.
Additionally, we can choose $N$ large enough in the construction such that
fewer than $N^3$ cuts have already been made in the game up to this stage. This
then allows Black to cut towards $y$, and then to cut arbitrarily at $A$ until
just a single string remains to a True entry, say to $x$.

Consider the middle coin of the $B$ wire. It has lost all of its strings to
$r$, and so its last string to $B$ can never be cut. This keeps $B$ attached
while Black cuts the rope from $B$ to $z$; at the other end, $z$ still has its
rope to $g$. The same argument lets Black cut the rope from $C$ to 0. This is
why we gave the ropes to $g$ width $N^4$: we choose $N$ large enough such that
even if White has spent every single turn cutting one of these ropes, it still
has a string left at this stage.

It remains to see who makes the last cut. In the wire joining $r$ to $A$, we
have a single string at $r$; the other two wires have lost their strings to
$r$, so each must keep one string to its row. Together with the string from
$A$ to $x$, the leaf string at $g$, and one single string on each of the four
paths ($x$, $y$, $z$, and 0), we count nine strings.

Now count the strings from the path ends to $g$. Recall that Black set $x$ to
True by cutting the string at $x$. Now the last string from $A$ holds $x$, and
the leaf holds $g$, so every string in the rope between $x$ and $g$ can be cut.
The other three ends that lost their strings are $z$, 1, and the unlabelled
False end of the $y$-path. Each must keep exactly one string to $g$. All four
ends that still have their strings to the middle coin in the path can lose
every string to $g$.

Thus, play will end with precisely $9+3=12$ strings still remaining. Had our
chosen entry been False instead of True (for example, if the last remaining
string from $A$ was to $y$ instead of $x$), then it would still have had its
string, and so we would end play with 13 strings instead of 12. We choose $N$
to be odd so that the board starts with an odd number of strings; we start with
eight single strings and 19 ropes, so an odd $N$ gives an odd total. Therefore,
by forcing a True entry, Black forces an odd number of cuts, hence making the
last move and winning. The same parity argument will work for our general
construction: there will always be an odd number of rows, each with an odd
number of entries, and a True entry will again allow one extra cut to $g$.

With a worked example under our belts, we turn now to the proof of the general
result: \cref{lem:coins-are-lava}. The full construction uses only the tricks
we have seen so far, just applied in a more general context.

\begin{proof}[Proof of \cref{lem:coins-are-lava}]
  Membership in \PSPACE{} is immediate since each move cuts one string, and
  legality of a cut is easily checked in polynomial time, so it remains to show
  hardness. Let $F$ be an instance of Schaefer's positive-CNF game. We may
  assume that $F$ has $c\ge1$ clauses, each containing at least one variable
  (otherwise the construction is trivial).

  For each clause with $k$ variables, we form a row by listing its variables
  and then appending $k-1$ copies of 1; this ensures that a majority of its
  $2k-1$ entries requires $k$ True entries. The copies of 1 supply $k-1$ of
  these required entries, and so the row is True precisely when at least one of
  its variables is True. Finally, we add $c-1$ rows that each consist of a
  single 0; this means that we have a majority of the resulting $2c-1$ rows
  only when we have $c$ True rows, which happens precisely when all $c$ clauses
  are satisfied.

  We now build the board just as in \cref{fig:cal-example}. We make a coin for
  each row, and we join it to the root $r$ by its own wire, of width $N$ at the
  root end and $N^2$ at the row end.

  We make one path of two single strings for each variable of $F$, labelling
  one end with the name of the variable. We make one further path with ends
  labelled 1 and 0, omitting the string at 1. The middle coin must keep its
  string to 0. This way, the 1 behaves as a variable already set to True, and
  0 as a variable already set to False.

  For each entry in a row, we join the row coin to the coin with that label by
  a rope of width $N^3$. Thus, two rows containing the same variable $x$ are
  both joined to the same coin labelled $x$, and two copies of 1 in a single
  row give two ropes from that row to the same coin labelled 1. Finally, we
  join both ends of every path to $g$ by ropes of width $N^4$, and give $g$ a
  private leaf. The resulting multigraph is connected, and it is
  straightforward to observe that it is also bipartite.

  To choose $N$, we count each rope once and each single string once, calling
  the total $t$. Take $N=2t+3$. We make this choice so that, while one player
  is cutting strings in ropes of width at most $w$, the other cannot cut
  through a whole rope of width $Nw$. Indeed, counting single strings as having
  width 1, there are at most $tw$ strings of width at most $w$ available to the
  first of these players. By the time they have finished, their opponent has
  had at most one more turn, so at most $2tw+1<Nw$ strings have been cut
  altogether. We will use this bound for widths 1, $N$, $N^2$, and $N^3$. Note
  also that $N$ is odd, and the board has at most $tN^4$ strings, which is
  polynomial in the size of the formula.

  Write $P$ for the player that has a winning strategy in the formula game, and
  $P'$ for their opponent. As play proceeds on the board, $P$ keeps a record of
  a play of the corresponding formula game. On each turn, $P$ claims the
  variables prescribed by their winning strategy on the formula game (setting
  True if $P$ is Black, and False if $P$ is White). If $P'$ sets a variable
  that is currently unclaimed in the formula game, then $P$ records that
  variable as the opponent's move; otherwise, if $P'$ plays elsewhere, then $P$
  records an arbitrary unclaimed variable as their opponent's move.

  In this way, every variable still unclaimed in the formula game is still
  unset on the board. Thus, $P$ can always make the move prescribed by their
  winning strategy on the formula game. Once all claims have been recorded, $P$
  finishes setting any variables that still happen to be unset on the board
  (which would appear if $P'$ plays elsewhere). If $P$ is Black, then they have
  set a True variable in every clause. If $P$ is White, then they have set
  every variable of some clause to be False. Hence, a majority of the rows have
  the value that $P$ desires.

  Consider the strings at the root $r$. Each row started with $N$ strings at
  the root end of its wire. Since a majority of the rows have the desired
  value, the wires towards those rows started with at least $N$ more strings
  between them than the remaining wires. Fewer than $N$ cuts have been made so
  far, and so those wires are still ahead. On each turn, $P$ can cut a string
  at $r$ towards a row of the wrong value, as long as one remains. Once all
  strings towards rows of the wrong value are gone, $P$ continues cutting at
  $r$ until only one string remains. These cuts are legal because each middle
  coin necessarily still has its wider rope of $N^2$ to its row. The last
  string at $r$ therefore leads towards a row of the desired value.

  Now $P$ will cut through the rope of width $N^2$ between the chosen row and
  the middle coin of its wire. The last string at $r$ holds the middle coin,
  and the ropes of width $N^3$ to the entries still hold the row, and so $P$
  can legally cut through the entire rope of $N^2$. Once this rope is gone, the
  row must still have a string to one of its entries. Here, we apply the same
  majority argument as before: each entry contributed $N^3$ strings, a majority
  of the entries have the desired value, and fewer than $N^3$ cuts have been
  made so far. Thus, by first cutting towards entries of the wrong value, $P$
  ensures that the last string at the chosen row leads to an entry of the
  desired value.

  Finally, $P$ cuts all remaining strings from the other rows to their entries.
  Each of these rows is held by the middle coin of its wire, which has lost all
  of its strings to $r$ already and therefore must keep at least one to its
  row. At the other ends, the entry coins still have their ropes to $g$, since
  fewer than $N^4$ cuts have been made so far.

  $P$ now plays arbitrarily until no legal cut remains. When play ends, each of
  the $2c-1$ wires has exactly one string left; the chosen wire keeps its
  string at $r$, and every other wire keeps one string at its row end. We also
  count a string from the chosen row to its chosen entry, and the string to the
  leaf at $g$.

  Suppose that $F$ has $n$ variables. For each variable, exactly one of its two
  end coins is still joined to its middle coin. Counting the string from 0 to
  its middle coin, too, gives $n+1$ strings. There are likewise $n+1$ end coins
  without a string to their middle coin: one for each variable, together with
  1. Each of these coins must keep one string to $g$ unless it is held by the
  chosen row. Every other string to $g$ can be cut, since the leaf holds $g$.

  If the chosen entry is False, then it is still joined to its middle coin.
  Thus, all $n+1$ of the end coins just counted must keep a string to $g$. This
  adds $2(n+1)$ strings to the $2c-1$ wire strings, the chosen row's string,
  and the leaf string, giving an odd total. If the chosen entry is True, then
  it is one of the $n+1$ coins that would otherwise need a string to $g$. The
  chosen row holds it, and so one fewer string remains, yielding an even total.

  The board started with an odd number of strings: each one of the $2c-1$ rows
  has an odd number of entries, and so there are an odd number of ropes from
  rows to entries; the wires contribute two ropes each, and $g$ has two ropes
  for each variable, one to 0, and one to 1. Thus, the total number of ropes is
  odd, and each has odd width. There are also $2n+2$ single strings: two for
  each variable, one joining 0 to its middle coin, and one joining $g$ to its
  leaf. Thus, a True entry gives an odd number of cuts, and a False entry gives
  an even number. Black moves first, and so $P$ makes the last cut in either
  case.
\end{proof}

\section*{Acknowledgements}
\addcontentsline{toc}{section}{Acknowledgements}

We thank \texttt{user30471} for the
\href{https://tex.stackexchange.com/a/424766/22649}{TikZ code on TeX Stack Exchange}
that we adapted for our Go board diagrams.

\bibliographystyle{plainurl}
\bibliography{bib}

\begin{thebibliography}{10}

\bibitem{LessonsInPlay:2007}
M.~H. Albert, R.~J. Nowakowski, and D.~Wolfe.
\newblock {\em Lessons in Play: An Introduction to Combinatorial Game Theory}.
\newblock A K Peters, Ltd., Wellesley, Massachusetts, 2007.

\bibitem{WinningWays:2001}
Elwyn~R. Berlekamp, John~H. Conway, and Richard~K. Guy.
\newblock {\em Winning Ways for Your Mathematical Plays}, volume~1.
\newblock A K Peters, Ltd., second edition, 2001.

\bibitem{buchin.ea:dots-and-boxes}
Kevin Buchin, Mart Hagedoorn, Irina Kostitsyna, and Max van Mulken.
\newblock {Dots \& Boxes} is {PSPACE}-complete.
\newblock In Filippo Bonchi and Simon~J. Puglisi, editors, {\em 46th
  International Symposium on Mathematical Foundations of Computer Science (MFCS
  2021)}, volume 202 of {\em Leibniz International Proceedings in Informatics
  (LIPIcs)}, pages 25:1--25:18. Schloss Dagstuhl -- Leibniz-Zentrum f{\"u}r
  Informatik, 2021.
\newblock \href {https://doi.org/10.4230/LIPIcs.MFCS.2021.25}
  {\path{doi:10.4230/LIPIcs.MFCS.2021.25}}.

\bibitem{2511.21888}
Kyle Burke, Caroline Cashman, Alfie Davies, Kanae Yoshiwatari, and Francesca
  Yu.
\newblock Mis\`{e}re {Partizan Arc Kayles} is {PSPACE}-complete, even on planar
  graphs, 2025.
\newblock \href {https://arxiv.org/abs/2511.21888} {\path{arXiv:2511.21888}}.

\bibitem{BurkeHearn2018}
Kyle Burke and Robert~A. Hearn.
\newblock {PSPACE}-complete two-color planar placement games.
\newblock {\em International Journal of Game Theory}, 48(2):393--410, 2019.
\newblock \href {https://doi.org/10.1007/s00182-018-0628-8}
  {\path{doi:10.1007/s00182-018-0628-8}}.

\bibitem{demaine.diomidova:strings-and-coins}
Erik~D. Demaine and Jenny Diomidova.
\newblock {Strings-and-Coins} and {Nimstring} are {PSPACE}-complete.
\newblock {\em Integers}, 21B:A7, 2021.
\newblock \href {https://doi.org/10.5281/zenodo.10939674}
  {\path{doi:10.5281/zenodo.10939674}}.

\bibitem{AlgGameTheoryGONC3}
Erik~D. Demaine and Robert~A. Hearn.
\newblock Playing games with algorithms: Algorithmic combinatorial game theory.
\newblock In Michael~H. Albert and Richard~J. Nowakowski, editors, {\em Games
  of No Chance 3}, volume~56 of {\em Mathematical Sciences Research Institute
  Publications}, pages 3--56. Cambridge University Press, 2009.
\newblock \href {https://doi.org/10.1017/CBO9780511807251.002}
  {\path{doi:10.1017/CBO9780511807251.002}}.

\bibitem{gallant.ea:edge-delete}
Robert~P. Gallant, Georg Gunther, Bert~L. Hartnell, and Douglas~F. Rall.
\newblock A game of edge removal on graphs.
\newblock {\em Journal of Combinatorial Mathematics and Combinatorial
  Computing}, 57:75--82, 2006.

\bibitem{DBLP:books/daglib/0023750}
Robert~A. Hearn and Erik~D. Demaine.
\newblock {\em Games, Puzzles, and Computation}.
\newblock A K Peters, Ltd., 2009.

\bibitem{huggan.nowakowski:conjoined}
Melissa Huggan and Richard~J. Nowakowski.
\newblock Conjoined games: {GO-CUT} and {SNO-GO}.
\newblock In Urban Larsson, editor, {\em Games of No Chance 5}, volume~70 of
  {\em Mathematical Sciences Research Institute Publications}, pages 395--402.
  Cambridge University Press, 2019.
\newblock \href {https://doi.org/10.1017/9781108641999.019}
  {\path{doi:10.1017/9781108641999.019}}.

\bibitem{johnson.ea:C123}
Matthew Johnson, Barnaby Martin, Jelle~J. Oostveen, Sukanya Pandey, Dani{\"e}l
  Paulusma, Siani Smith, and Erik~Jan van Leeuwen.
\newblock Complexity framework for forbidden subgraphs {I}: The framework.
\newblock {\em Algorithmica}, 87(3):429--464, 2025.
\newblock \href {https://doi.org/10.1007/s00453-024-01289-2}
  {\path{doi:10.1007/s00453-024-01289-2}}.

\bibitem{moore:anti-atari-go}
John Moore.
\newblock {Anti Atari Go}.
\newblock Sensei's Library, version 1, 1 April, 2005.
\newblock \url{https://senseis.xmp.net/?AntiAtariGo:v1}.

\bibitem{DBLP:journals/jcss/Schaefer78}
Thomas~J. Schaefer.
\newblock On the complexity of some two-person perfect-information games.
\newblock {\em Journal of Computer and System Sciences}, 16(2):185--225, 1978.
\newblock \href {https://doi.org/10.1016/0022-0000(78)90045-4}
  {\path{doi:10.1016/0022-0000(78)90045-4}}.

\bibitem{SiegelCGT:2013}
Aaron~N. Siegel.
\newblock {\em Combinatorial Game Theory}, volume 146 of {\em Graduate Studies
  in Mathematics}.
\newblock American Mathematical Society, 2013.
\newblock \href {https://doi.org/10.1090/gsm/146} {\path{doi:10.1090/gsm/146}}.

\end{thebibliography}

\end{document}